\documentclass[nofootinbib,
pra,
twocolumn,
amsmath,amsfonts,
superscriptaddress,
]{revtex4-2}
\usepackage{enumitem}
\usepackage[T1]{fontenc}
\usepackage{hyperref}
\usepackage{amssymb}
\usepackage{graphicx}
\usepackage{bbm}
\usepackage{grffile}
\usepackage{physics}
\newcommand{\param} {{\Phi}}
\newcommand{\paramOpt} {\param_{\rm opt}}
\newcommand{\tOpt} {t_{\rm opt}}
\newcommand{\im} {{\iota}}
\newcommand{\dT}[1] {{\partial}_{#1}}

\newcommand{\qfi} {\mathcal{F}_\param}
\newcommand{\cfi} {F_C(\param)}
\newcommand{\cfiEP} {F_{\rm EP}(\param)}
\newcommand{\Par}{\hat{\mathcal P}}
\newcommand{\Jz}{\hat J_{\rm z}}
\newcommand{\Jx}{\hat J_{\rm x}}
\newcommand{\Jy}{\hat J_{\rm y}}
\newcommand{\ketJ}[1] {\ket{J,#1}}
\newcommand{\braJ}[1] {\bra{J,#1}}
\newcommand{\catTh}[1] {\ket{\Psi(#1)}}
\newcommand{\mAv}{\overline{m}}
\newcommand{\mVar}{\sigma^2}
\newcommand{\x}[1] {{\sigma^{\rm x}_{#1}}}
\newcommand{\z}[1] {{\sigma^{\rm z}_{#1}}}
\newcommand{\normN}[1] {{\mathcal{A}}_{#1}}
\renewcommand{\eqref}[1]{Eq.~(\ref{#1})}
\renewcommand{\[}{\begin{equation}}
\renewcommand{\]}{\end{equation}}

\begin{document}
\title{Robustness of spin state superpositions for noisy quantum metrology}
\author{Gabriela Wójtowicz}
\email{gabriela.wojtowicz@uni-ulm.de}
\affiliation{Institut f\"ur Theoretische Physik und IQST, Albert-Einstein-Allee 11, Universit\"at Ulm, D-89081 Ulm, Germany}
\thanks{These authors contributed equally.}
\author{Trinidad B. Lantaño}
\email{trinidad.lantano-pinto@uni-ulm.de}
\affiliation{Institut f\"ur Theoretische Physik und IQST, Albert-Einstein-Allee 11, Universit\"at Ulm, D-89081 Ulm, Germany}%
\thanks{These authors contributed equally.}
\author{Susana F. Huelga}
\affiliation{Institut f\"ur Theoretische Physik und IQST, Albert-Einstein-Allee 11, Universit\"at Ulm, D-89081 Ulm, Germany}%
\author{Martin B. Plenio}
\affiliation{Institut f\"ur Theoretische Physik und IQST, Albert-Einstein-Allee 11, Universit\"at Ulm, D-89081 Ulm, Germany}%
\begin{abstract}
Quantum metrology faces major challenges in noisy environments, where decoherence rapidly degrades useful quantum resources. 
We investigate the dynamics of the precision limits given by the quantum Fisher information (QFI) for phase estimation under spatially correlated dephasing.
We characterize the dynamics of the QFI by the sensitivity and degradation indicators that can be obtained as analytical expressions derived using perturbative theory treatment. 
These short-time and weak-noise formulas yield analytic insight into how collective-spin moments govern both (i) the noiseless sensitivity and (ii) the leading noise-induced degradation of metrological usefulness. 
We identify a trade-off that is intrinsic to our commuting encoding-noise structure. 
We analyze the QFI dynamics for the Gaussian spin state (GSS) superpositions, encompassing spin coherent state (SCS), Dicke state superpositions, spin-squeezed states, and GHZ-like states. 
Predictions from indicators of the QFI dynamics are compared to both the quantum Cram\'er--Rao bound and the measurement-specific sensitivity bounds for an optimal parameter and interrogation time under a finite total time resource.
When possible, we analytically derive the measurement-specific sensitivity bounds for spin-projection and parity-based measurements.
\end{abstract}

\maketitle

Phase estimation is a paradigmatic problem in quantum metrology, where quantum correlations can enhance the precision of parameter estimation beyond classical limits~\cite{caves_quantum-mechanical_1981}
In practice, however, decoherence degrades quantum resources, and the resulting precision depends critically on the type and structure of the noise~\cite{huelga_improvement_1997,HaaseHuelga}.
Among the possible noise channels, the most detrimental are those whose jump operators commute with the encoding Hamiltonian: even error correction protocols cannot restore Heisenberg scaling in this case, and a generic noisy estimation protocol is effectively reduced to a residual regime with such commuting dephasing~\cite{chin_quantum_2012,smirne_huelga_2016,demkowicz_adaptive_2017,haase_fundamental_2018,zhou_achieving_2018,zhou_optimal_2020}.
Understanding this fundamental bottleneck motivates the present work, where we develop a perturbative framework to systematically characterize the metrological performance of symmetric spin states under commuting dephasing noise.

\section{Noisy quantum metrology}
\label{sec:noisy_metrology-overview}

In quantum metrology, the goal is to estimate an unknown parameter $\param$, by first encoding it in the state of a quantum system and then performing measurements. The results of these measurements carry statistical information about $\param$, and the precision of the estimation depends on how sensitively the probability of the outcome changes in response to small variations of the parameter~\cite{braunstein_statistical_1994,giovannetti_quantum_2006, Giovannetti:2011chh, Tóth_2014,HaaseHuelga}.\\
In the limit of large measurement statistics, the precision of any unbiased estimator is established by the celebrated Cram\'er--Rao bound~\cite{cramer1999mathematical}
\begin{equation}\label{eq:cramer_rao}
\Delta^2\tilde \param \geq \frac{1}{\cfi \left[P(\vec{x}|\Phi) \right]} ,
\end{equation}
where $\Delta^2 \tilde{\param}$ denotes the variance of the parameter's statistical estimator, and $\cfi \left[P(\vec{x}|\Phi)\right]$ is the classical Fisher information (FI) associated with the conditional probability distribution $P(\vec{x}|\param)$ of outcomes $\vec{x}$ given the parameter $\param$. This distribution, often referred to as the likelihood function, encodes the dependence of measurement outcomes on the parameter to be estimated.\\
In quantum systems, outcome probabilities depend on the choice of measurement, since the statistics are determined by the specific observable being probed. For a quantum measurement described by a positive operator valued measure (POVM) element $\hat{E}_{\vec{x}}$, the probability of obtaining outcome $\vec{x}$ is given by Born's rule: $P(\hat{E}_{\vec{x}} | \Phi)=\Tr{\hat{E}_{\vec{x}} \hat{\rho}(\Phi)}$, where $\hat{\rho}(\Phi)$ is the quantum state parametrized by $\Phi$. Different choices of measurement operators $\hat{E}_{\vec{x}}$ yield different likelihood functions, and thus different FI.\\
The quantum Fisher information (QFI) is defined as the maximum of the classical Fisher information over all possible quantum measurements. It quantifies the best precision achievable in estimating a parameter $\Phi$ encoded in a state $\rho(\param)$ through the optimal measurement. The achievable precision is bounded by the quantum Cram\'er--Rao bound $
\Delta \tilde{\param} \,\geq\, \qfi^{-1/2}$,
where the QFI $\qfi$ depends both on the probe state and on the encoding of $\param$. Operationally, the QFI measures the statistical distinguishability, or quantum statistical speed, of the parameterized states~\cite{braunstein_statistical_1994,Smerzi_StatisticalSpeedQFI}.\\ 
Throughout this work, we use ``ultimate precision'' to refer to the Cram\'er--Rao bound of the estimation strategy under discussion \emph{after} optimization over both the encoded parameter and the interrogation time at fixed total time $T$. The term therefore denotes the end point of the optimization, not optimality over all measurements: the readout it refers to -- spin projection, parity, or the optimal measurement saturating the QFI -- is stated explicitly in each case.
\\
Beyond metrological precision, the QFI also serves as an entanglement witness. A collective rotation $e^{i \param\hat{J}_{\hat{n}}}\rho e^{-i \param \hat{J}_{\hat{n}}}$ can be used to certify multipartite entanglement: an $N$-particle state $\rho(\param)$ must be at least $\lfloor \qfi/N \rfloor$-partite entangled, where $\lfloor x \rfloor$ denotes the integer part of $x$ \cite{pezzeQFIEntangl}.\\
For separable states, the so-called standard quantum limit (SQL) bounds the precision as $\qfi \leq N$ \cite{PhysRevLett.102.100401}.
By contrast, maximally entangled states such as GHZ states \cite{Greenberger1989} can reach the Heisenberg limit (HL), achieving $\qfi \sim N^2$. 
However, under local dephasing noise, the metrological advantage of GHZ states is lost: their precision becomes equivalent to that of separable states \cite{huelga_improvement_1997,HaaseHuelga}.\\
{While it is well established that the precision scaling in quantum metrology is highly sensitive to noise, the exact impact depends strongly on the type, and in some cases it is possible to partially recover a quantum advantage.}
For local Markovian dephasing, the asymptotic scaling with the number of particles is limited to the SQL, where the bound on the estimation precision scales as $\Delta \tilde{\param} \sim 1/\sqrt{N}$, regardless of the state being separable or entangled \cite{huelga_improvement_1997,Escher:2011eff,Demkowicz-Dobrzanski:2012dnq}. However, relaxing the Markovian assumption and introducing temporally correlated noise yields new fundamental limits~\cite{chin_quantum_2012, PhysRevA.98.020102}: in particular, for GHZ states a scaling of $\Delta \tilde{\param} \sim N^{-3/4}$ was predicted~\cite{smirne_huelga_2016}.
\\
When atoms are spatially close, as in an optical lattice or an ion trap, local dephasing is no longer the dominant source of noise. Instead, the main concern is \textit{spatially correlated} dephasing. Several experiments have confirmed that this type of dephasing is the major source of decoherence in ion traps \cite{doi:10.1126/science.1057357,PhysRevLett.95.060502,Roos:2006tra,PhysRevLett.106.130506}.\\ 
Remarkably, certain entangled states can be engineered to remain entirely unaffected by correlated dephasing. These states live in a decoherence-free subspace (DFS) and have been experimentally realized, providing exceptional enhancements for quantum metrology with trapped ions \cite{doi:10.1126/science.1057357,Roos:2006tra}.
{However, when noise and encoding share the same operator, the DFS coincides with the eigenspace of the encoding Hamiltonian, so that no phase can be imprinted.
States outside the DFS, in turn, exhibit \textit{superdecoherence}~\cite{doi:10.1098/rspa.1996.0029}, an effect arising from collective entanglement between the atomic ensemble and the environment without any exchange of energy. Experimentally, measurements of the dephasing rate of GHZ states have revealed a quadratic scaling with the number of atoms~\cite{PhysRevLett.106.130506}.}
\\
{While in this work we focus on perfectly correlated dephasing, a more general formalism in which correlations decay over a finite correlation length was derived in \cite{PhysRevA.87.052138}.} Building on this, the metrological performance of $N$ hydrogen-like ions was studied in \cite{Jeske_2014}, showing that certain entangled states can preserve Heisenberg scaling $\Delta \tilde{\param} \sim 1/N$ even when the correlation length is small.\\
{Although the metrological performance of atomic systems under both correlated and local noise has been extensively studied, comparatively little attention has been paid to the structural features that make a quantum state robust for metrology under these decoherence mechanisms.}
In \cite{huelga_improvement_1997}, it was shown that probe states exhibiting symmetry under particle exchange are optimal for quantum phase estimation in the presence of local decoherence. This same family of states was later demonstrated to yield the maximal QFI under spatially correlated noise \cite{dorner_quantum_2012}.\\
Despite these insights, the class of symmetric states remains remarkably broad, encompassing not only highly entangled states but also separable ones.
This raises a key question: {what} intrinsic features of symmetric states determine their robustness in noisy metrological scenarios?
{While fast control protocols can restore Heisenberg scaling in certain noisy settings~\cite{Dur2014improvedquantummetrology,Sekatski2017quantummetrology}, thereby evading or modifying fundamental trade-offs, understanding the passive resilience of states without active control remains important for identifying inherent state properties that favor metrological robustness.}\\
{To address this, we focus on spatially correlated dephasing and study a system of $N$ spins symmetric under particle exchange, described within the fully symmetric subspace of fixed angular momentum $J=N/2$. This representation enables us to characterize such states via their probability distribution on the Bloch sphere and identify how moments of these distributions govern metrological performance under dephasing.} \\
We begin by deriving a perturbative expression for the QFI of a pure state undergoing weak spatially correlated dephasing noise.
The resulting formula decomposes the QFI into a leading noiseless term, which quantifies the sensitivity to the encoding, and a linear correction that captures the noise-induced degradation.
This perturbative decomposition provides analytical insight into the behavior of symmetric states under correlated dephasing, making explicit a key trade-off: the same collective-spin moments that enhance unitary sensitivity also increase the leading noise-induced decay of the QFI. Although this trade-off is well known for particular states such as the GHZ state, our formulas quantify it systematically across a broad class of probes.\\
{The simplicity of the obtained expressions enables closed-form results for a broad class of relevant probes, including superpositions of spin-squeezed states~\cite{kitagawa_squeezed_1993,ma_quantum_2011}, spin coherent states (SCSs)~\cite{kitagawa_squeezed_1993}, Dicke states~\cite{dicke1954coherence,prevedel2009experimental,lucke2014detecting,Tóth_2014}, and more generally, Gaussian spin states (GSSs)~\cite{huang_optimized_2008,pezze_quantum_2021}.\\
We first analyze squeezed spin states~\cite{kitagawa_squeezed_1993,ma_quantum_2011} in the moderate squeezing regime. In this setting, we derive closed-form expressions for the leading noiseless and noise-induced contributions to the QFI. Here, we show that in the regime considered, the sensitivity gained through squeezing is not outweighed by the increased susceptibility to noise. While the effect of dephasing on squeezed states has been studied previously, mostly in the context of specific protocols such as Ramsey spectroscopy~\cite{UlamOrgikh2001}, explicit QFI-based analyses remain scarce; one notable exception is Ref.~\cite{Zhong2014}, which considers squeezed states under collective dephasing.  More recently, optimal phase estimation protocols under correlated dephasing were benchmarked against fundamental bounds, showing that spin squeezed probes are practically optimal for positively correlated phase fluctuations, while more general tensor network optimized strategies can provide advantages in the presence of negatively correlated noise~\cite{ghosh_optimal_2025}.\\
We then turn to superpositions of spin coherent states (SCCSs). For this family, we derive closed-form expressions for the leading noiseless and noise-induced terms at arbitrary orientation. These results provide an analytical explanation for the enhanced robustness of SCCSs compared with GHZ states~\cite{huang_quantum_2015}. For superpositions of Dicke states, which arise as the zero-width limit of the GSSs, we derive an exact closed-form expression for the QFI under spatially correlated dephasing. We show that, at fixed orientation, their leading large-$J$ noiseless and noise-induced contributions coincide with those of SCCSs, establishing a close analytical correspondence between the two families. However, despite this common short-time scaling, Dicke-state superpositions suffer an exponential loss of coherence and are therefore less robust at finite interrogation times than SCCSs.\\
We also study imperfect GHZ-state preparations, modeled as superpositions of extremal GSSs with non-zero magnetization width, \textit{i.e.} $\sigma \ne $0. We show that these broadened GHZ-like states can preserve metrological enhancement -- given by the QFI -- over longer interrogation times than perfectly prepared GHZ states.\\
Beyond the QFI, we examine experimentally motivated measurement schemes, focusing on spin-projection and parity measurements. For squeezed states with spin-projection readout, we find that the estimation precision improves with increasing squeezing, although the ultimate noise-limited bound $\gamma/T$ cannot be surpassed.\\
For superpositions of Dicke states, the parity measurement saturates the quantum Cram\'er--Rao bound and therefore realizes the optimal measurement for this family. Optimizing over the encoded parameter and interrogation time, we find that the achievable precision at arbitrary orientation is bounded from below by $e\gamma/T$, generalizing the result previously known for the GHZ state~\cite{huelga_improvement_1997} to the full family of Dicke-state superpositions.\\
For GHZ-like states, however, the QFI advantage associated with finite broadening is not captured by parity readout. Instead, the performance of this measurement degrades because parity is tailored to the ideal GHZ ansatz with zero broadening. Extracting the advantage of broadened GHZ-like states therefore requires a measurement strategy adapted to their finite width $\sigma$.\\
Finally, we optimize the Cram\'er--Rao bound over both the encoded parameter and the interrogation time to compare the achievable precision of the GHZ state and the SCS under correlated dephasing. In this comparison, we use the measurement that is optimal in the noiseless case: parity for the GHZ state and spin projection for the SCS. Both states yield the same asymptotic scaling with total time and noise amplitude, although the GHZ state exhibits a \textit{larger} estimation error by a factor of $e$. For the SCS, the estimation error approaches the noise-limited floor $\gamma/T$, with a sub-leading correction that vanishes as $\mathcal{O}(N^{-2/3})$ with system size.\\
The analytical results are complemented by numerical simulations.}

\section{Model}
\label{sec:model}

We consider a scenario in which an ensemble of spins or two-level atoms is used to estimate the amplitude of a magnetic field in the presence of noise. 
Dephasing noise in atomic ensembles often reflects the averaged effect of fluctuations in stray electromagnetic fields. It can be modeled as all particles experiencing random magnetic field fluctuations that induce stochastic energy shifts in their internal transitions (see \cite{dorner_quantum_2012} for a detailed derivation). An analogous description applies to two‑mode Bose–Einstein condensates (BECs) \cite{huang_quantum_2015}.
In most cases, these fluctuations lack temporal correlations and can therefore be treated as Markovian, while their spatial correlations depend on the scale of the system. When the spatial extent of an atomic ensemble is much smaller than the correlation length of the stray field, the dominant noise is modeled by spatially correlated dephasing \cite{doi:10.1126/science.1057357,PhysRevLett.95.060502,Roos:2006tra,PhysRevLett.106.130506}, described by
\begin{align}
\label{eq:master_eq_correlated}
\dT{t} \rho  &= -i\param[\Jz, \rho] + \gamma \mathcal{D} \left[\rho\right] , 
\end{align}
where 
$\param$ is the parameter to estimate, and $\gamma$ the dephasing strength, note that here we take $\param$ to have units of frequency, so that the noiseless Hamiltonian is $\hat{H}= \hbar \param \Jz$. The dephasing dynamics reads
\[
\mathcal{D} \left[\rho\right]= \Jz\rho\Jz-\frac{1}{2}\{\Jz^2, \rho\}  .
\]
Note that with this evolution, the off-diagonal density matrix elements in the $\Jz$ eigenbasis decay as $\exp{-\gamma t(m-n)^2/2}$ (see App.~\ref{app:methods}) rapidly suppressing the coherences that serve as the resource for sensing rotations generated by $\Jz$. 

\section{Optimal sensitivity and robustness}
\label{sec:qfi_perturbative}

Determining the sensitivity set by the QFI generally demands a full eigendecomposition of the density matrix.
In generic many-body settings this step is computationally demanding, since the dimension of the Hilbert space grows exponentially with the number of particles.
In the present problem, however, permutation symmetry confines the dynamics to the $(2J+1)$-dimensional symmetric subspace, so that exact diagonalization is explicitly feasible, and we use it for all numerical results reported below.
What remains difficult is analytical rather than numerical: even for pure states the eigendecomposition of the evolved density matrix does not yield closed-form expressions, and therefore offers no transparent link between the structure of a probe state and its metrological performance.
To expose that link, we characterize the essential behavior of the state through quantifiers derived using a perturbative theory approach. 
In the short-time and weak noise regime, these quantities are directly accessible as the zeroth- and first-order corrections to the QFI, providing powerful analytical insight without the need for full diagonalization.
\\
In the limit of weak noise and short interrogation time, that is, $\gamma t \norm{\mathcal{D} \left[\rho_0\right]}\ll 1$ in~\eqref{eq:master_eq_correlated}, the QFI is approximated by 
\begin{align}
\label{eq:qfi_perturbative}
    \qfi[\rho_0] \approx \qfi^{(0)}[\rho_0]  - \gamma t \cdot \qfi^{(1)}[\rho_0] , 
\end{align}
where $t$ is the interrogation time. 
While the expression in~\eqref{eq:qfi_perturbative} yields a reliable approximation of the QFI only for small perturbations, it already reflects the QFI dynamics. 
The leading term $\qfi^{(0)}$ captures the ideal, noiseless scaling, directly measuring how sharply the state responds to the encoding, while the subleading {$\gamma t$}-dependent correction $\qfi^{(1)}$ reveals exactly how fast the QFI is eroded by noise. 
To reflect these properties, we refer to the quantity $\qfi^{(0)}$ as the {\it sensitivity indicator} and $\qfi^{(1)}$ as the {\it degradation indicator}. 
Ideally, we aim for maximum sensitivity and minimal degradation indicators, meaning strong sensitivity to the encoding and weak degradation due to the noise. 
Crucially, the indicators can be derived as closed-form analytical expressions, offering a clear and powerful characterization of the metrological performance of the probe state in a noisy scenario.
For an extended comment on the perturbative treatment for the QFI, see App.~\ref{app:qfi_perturbative_dephasing}. 
\\
For the noisy model introduced in Sec.~\ref{sec:model}, the expression for the sensitivity indicator reads
\[\label{eq:F0_correlated}
\qfi^{(0)}[\Psi] = 4 \nu t^2 \Delta^2 \Jz ,
\]
where the variance $\Delta^2 \Jz=\langle \Jz^2 \rangle - \langle \Jz\rangle^2$ takes the expectation values at the initial state, that is, $\langle\hat O \rangle=  \bra{\Psi}\hat O\ket{\Psi}$ for the pure input state $\ket{\Psi}$ and operator $\hat O$. 
The degradation indicator
due to spatially correlated noise reads
\begin{align}\label{eq:F1_correlated_general}
     \qfi^{(1)}[\Psi] =4 \nu t^2 (\Delta^2 \Jz) \left[
4(\Delta^2 \Jz)
+\frac{\mu_3^2}{(\Delta^2 \Jz)^2}
\right],
\end{align}
where $\mu_3= \langle \left(\Jz - \langle\Jz \rangle \right)^3\rangle$. Hence, the degradation indicator depends on the second and third central moments of $\Jz$. 
For the derivation, see Appendix~\ref{app:qfi_perturbative_dephasing}.
\\
For states symmetric under reflection across the $xy$-plane, \textit{i.e.}, under $\Jz \rightarrow -\Jz$, all odd moments of $\Jz$ vanish. In particular, $\langle \Jz \rangle = 0$ and $\mu_3=0$, so that the degradation indicator reduces to
\begin{align}\label{eq:F1_correlated}
     \qfi^{(1)}[\Psi] =16 \nu t^2 (\Delta^2 \Jz)^2.
\end{align}
Note that $\mu_3$ enters~\eqref{eq:F1_correlated_general} only through $\mu_3^2\geq0$. At fixed sensitivity $\qfi^{(0)}$, that is, at fixed $\Delta^2\Jz$, the degradation indicator is therefore minimized precisely by those states with $\mu_3=0$. Mirror-symmetric probes are in this sense the least fragile ones at a given noiseless sensitivity, which motivates the families considered in Sec.~\ref{sec:example}.
\\
Since both $\qfi^{(0)}$ and $\qfi^{(1)}$ depend on the same moment of the encoding and jump operator $\Jz$, the perturbative formulas imply a trade-off between the sensitivity and degradation due to noise.
In particular, states with a large variance $\Delta^2 \Jz$ offer high sensitivity in the absence of noise, but at the same time, they are inevitably more susceptible to decoherence.

\section{Example}
\label{sec:example}

In this section, we dive into several key families of quantum probes that frequently appear in the literature. 
Each of the presented families of states fits within the unifying framework of the Gaussian spin state (GSS) ansatz.
For each of them, we will characterize the QFI dynamics by calculating analytically the sensitivity and degradation indicators in Sec.~\ref{sec:qfi_perturbative}. 
The indicators characterize the quantum Cram\'er--Rao bound, that is, the precision attainable with the measurement that saturates the QFI. \\
Later, in Sec.~\ref{sec:optimize}, we compare this bound with the Cram\'er--Rao bound obtained from specific, practically motivated measurements, such as spin projection or parity measurements.
This measurement-specific bound is determined by the corresponding classical error-propagation bound. 
The comparison allows us to identify the conditions under which a given measurement can approach the optimal precision for a fixed interrogation time.\\
Whenever possible, we derive analytical expressions for the classical and quantum Cram\'er--Rao bounds, otherwise, we compute the QFI numerically. 
For correlated noise, the dynamics remain in the fully symmetric subspace, allowing efficient evaluation via eigendecomposition.
For the comment about the numerical methods, see App.~\ref{app:methods}
\\
{The results we present in the following operate at four distinct levels of rigor, which we summarize here for the reader's orientation. \emph{Exact} closed-form expressions are obtained for Dicke-state superpositions. \emph{Perturbative} expressions valid to leading order in $\gamma t$, in the regime $\gamma t\,\lVert\mathcal{D}(\rho)\rVert \ll 1$, govern the sensitivity and robustness measures and the GHZ-like Cram\'er--Rao bounds. \emph{Asymptotic} expressions valid for $J\gg 1$ are derived for spin coherent and squeezed states as well as GHZ-like states. Beyond these analytical regimes, several trends are established \emph{numerically} via exact diagonalization in the symmetric subspace. We indicate the applicable regime when stating each result.}
\\
In Sec.~\ref{sec:gss}, we introduce the GSS ansatz and analyze its properties within the framework of quantum metrology. 
In particular, we employ the sensitivity and robustness quantifiers derived in Sec.~\ref{sec:qfi_perturbative} to perform a comparative assessment of the probe ansätze across the relevant parameter space.
After this broad overview, we now turn to concrete families of probe states. 
In Sec.~\ref{sec:sss}, we investigate the properties of the spin coherent state and the impact of squeezing. 
In Sec.~\ref{sec:scs_cat}, we analyze the symmetric superposition of two classical states, that is, spin coherent states.
In Sec.~\ref{sec:dicke_cat}, we investigate the case in which the probe is initialized in a coherent superposition of Dicke states. 
In Sec.~\ref{sec:ghz_like}, we investigate a class of states called GHZ-like states. 
\\
Having developed intuition for the QFI dynamics, we optimize the sensing experiment under a time constraint. 
In Sec.~\ref{sec:optimize}, we derive the precision bound from the QFI and compare it with practically motivated spin-projection and parity measurements. 
For the spin-projection measurement in Sec.~\ref{sec:optimize_Jy}, we analyze the spin-squeezed states. 
We show that squeezing initially enhances estimation precision but eventually degrades it once the state is over-squeezed. 
For parity measurement in Sec.~\ref{sec:optimize_Par}, we analyze the superposition of Dicke states and GHZ-like states. 
For Dicke-state superposition and GHZ states, the quantum Cramér–Rao bound is saturated by the parity measurement. 
For GHZ-like states, the parity measurement is not optimal but remains practically motivated for states sufficiently close to the GHZ. 
We show that endowing the superposition with a finite variance significantly improves its robustness against noise compared to the ideal GHZ state.

\subsection{Gaussian spin state ansatz}
\label{sec:gss}
\begin{figure}[h!]
    \centering
    \includegraphics[width=0.7\linewidth]{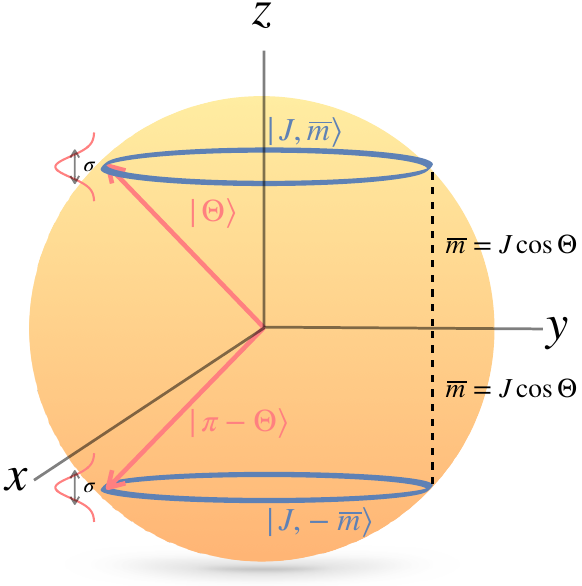}
    \caption{\textbf{Bloch sphere representation of the Gaussian spin states in \eqref{eq:probestate_definition}}. Each component of the superposition is shown as a pink arrow, with its projection onto the $z$-axis represented by a Gaussian distribution of width $\sigma$. Pairs of Dicke states $\ket{J,\overline{m}}$ and $\ket{J,-\overline{m}}$, which share the same $z$ coordinate, are highlighted by blue rings.
    }
    \label{fig:BlochSphere}
\end{figure}

We work in the fully symmetric subspace of $N$ identical spin-$1/2$ particles.
This is the natural setting for the collective metrological model considered in Sec.~\ref{sec:model}, as both the parameter encoding and the correlated dephasing channel are generated by collective spin operators, and therefore preserve permutation symmetry.
An initially symmetric probe thus remains within the spin-$J$ sector with $J=N/2$, where it can be expanded in the Dicke basis as
\begin{align}\label{eq:stateexpansion}
\ket{\Psi} = \sum_{m=-J}^J c_m \ketJ{m} ,    
\end{align}
where $\ketJ{m}$ denotes the Dicke state~\cite{DICKE1} with total spin $J$ and magnetization $m$ and $c_m$ is the corresponding state amplitude.
The Dicke state $\ketJ{m}$ is an equal-weighted superposition of all possible spin configurations in an ensemble of $N = 2J$ spins-$1/2$ with a total magnetization $m$. 
Within the symmetric subspace, the state can be represented on the spin-$J$ Bloch sphere in Fig.~\ref{fig:BlochSphere}. 
The coefficients $c_m$, thus, encode the geometric distribution of the state on the Bloch sphere. \\
In this paper, we set the state ansatz to be a superposition of Gaussian spin states, that is, 
\begin{align} \label{eq:probestate_definition}
\catTh{\Theta, \sigma} = \normN{\catTh{\Theta, \sigma}} \left(\ket{\Theta, \sigma}+\ket{\pi-\Theta, \sigma}\right) ,
\end{align}
where $0 \leq \Theta \leq \pi/2$ is the polar angle, $\normN{\catTh{\Theta, \sigma}}$ is a normalization factor, and
\begin{align}\label{eq:GSS}
\ket{\Theta, \sigma} = \sum_{m=-J}^J c_m(\Theta, \sigma) \ketJ{m}
\end{align}
is a {\it Gaussian spin state} (GSS)~\cite{imamoglu_inhibition_1997,pezze_quantum_2018,alexander_generating_2020}.
The GSS is defined by the Gaussian distribution over the Dicke basis, with its coefficients described by the corresponding (non-normalized) probability distribution
\begin{align}\label{eq:coeffsbinom}
|c_m(\Theta, \sigma)|^2 = \exp{-\frac{(m-\mAv)^2}{2\mVar}} ,
\end{align}
where $\mAv=J\cos\Theta$ is the mean, fixed by the angle $\Theta$, and $\sigma$ is the standard deviation (or width) of the distribution. 
In this paper, we choose the coefficients $c_m(\Theta, \sigma)$ to be real-valued in such a way that the resulting GSS superposition state exhibits mirror symmetry, as illustrated in Fig.~\ref{fig:BlochSphere}.

Characterizing our probe state by the ansatz in~\eqref{eq:coeffsbinom} allows us to develop an understanding of the impact of the orientation of the spin state and its variance on the robustness of metrological precision. 
Physically, the GSS serves as an approach to approximate squeezed states~\cite{imamoglu_inhibition_1997,pezze_quantum_2018,alexander_generating_2020} where the variance of the distribution can be smoothly controlled. 
It can also be derived either from the action of a Kraus operator via tailored squeezing protocol~\cite{imamoglu_inhibition_1997,huang_optimized_2008} or with a Gaussian distribution over the state components~\cite{pezze_quantum_2021}.
The GSS superposition is a unifying framework that encompasses several metrologically relevant families of spin states, such as spin coherent state (SCS) ~\cite{
PhysRevA.56.2249,PhysRevA.90.045804,huang_quantum_2015,PhysRevB.103.L100403} and Dicke state \cite{Tóth_2014,zhang_quantum_2014,opatrny_counterdiabatic_2016,carrasco_extreme_2022,carrasco_dicke_2024} superpositions, GHZ-like states~\cite{zhang_fast_2024,kielinski_ghz_2024} and spin-squeezed states~\cite{kitagawa_squeezed_1993}.\\
In the limit of large $J$ the binomial distribution of the spin coherent state is well approximated by Gaussian. 
Setting the variance $\mVar=J/2\sin^2\Theta$, the expression in~\eqref{eq:probestate_definition} becomes a superposition of two SCSs. 
The SCS superposition can be prepared by adiabatic evolution in atomic Bose--Einstein condensates~\cite{PhysRevLett.97.150402,doi:10.1126/science.aag1106} or by performing sequential measurements on a central spin system~\cite{Lantano2025unlockingheisenberg}. 
The metrological properties of such states in the presence of dephasing noise have been thoroughly analyzed by Huang et al.~\cite{huang_quantum_2015, PhysRevA.98.012129}, who demonstrated that they yield enhanced robustness of estimation precision under noisy conditions when compared to GHZ states.
\\
The superposition of Dicke states is recovered in the limit of vanishing variance, \textit{i.e.}, $\sigma \to 0$, of the GSS ansatz in~\eqref{eq:probestate_definition}. 
Dicke states are known for their strong entanglement and have been demonstrated to be optimal for unitary metrology~\cite{Tóth_2014,zhang_quantum_2014}. 
From an experimental perspective, this class of states can be interpreted as the limit of extremely squeezed states~\cite{sorensen_entanglement_2001,opatrny_counterdiabatic_2016,carrasco_extreme_2022,carrasco_dicke_2024}.
A single Dicke state, being a steady state of the master equation in~\eqref{eq:master_eq_correlated}, cannot be used for the encoding in Sec.~\ref{sec:model}. However, a superposition such as a GHZ state can saturate the Heisenberg limit. 
\\
Finally, the GHZ-like states are obtained from the probe ansatz in~\eqref{eq:probestate_definition} by taking $\Theta=0$ and $\sigma\geq0$. 
In the limit of vanishing variance, we recover the celebrated GHZ state. 
Yet, even away from this ideal case, the system reveals striking features in the so‑called GHZ‑like regime, where an imperfect GHZ state still exhibits rich and intriguing properties.

\subsection{QFI dynamics and impact of noise}

To assess the effectiveness of the GSS ansatz in noisy phase estimation, we compute the corresponding sensitivity and degradation indicators in~\eqref{eq:F0_correlated} and~\eqref{eq:F1_correlated}, respectively. 
In a short time, the scaling can be explained using the perturbative approximation of the QFI, that is, $\qfi\approx\qfi^{(0)} - \gamma t \qfi^{(1)}$. 
The validity of the perturbative QFI expansion relies on the assumption that noise-induced decoherence remains weak, \textit{i.e.} $\gamma t (m-n)^2/2 \ll 1$ for those coherences $\rho_{mn}$ with significant weight $c_m, c_n$, see App.~\ref{app:qfi_perturbative_dephasing}, 
so the relevant scale is set by the coherence spread with the index separation $|m-n|$. 
For example, for SCCSs with $\Theta=\pi/2$ the dominant contributions of the state are within $-\sqrt{J} \lesssim m\lesssim\sqrt J$, implying that the typical separation is $|m-n| \sim \sqrt{J}$. 
This yields the mild perturbative condition $\gamma t \ll 1/J$. 
By contrast, for GHZ states, the coherences are between the extremal Dicke components with magnetic quantum numbers $m = \pm J$, which are separated by $|m-n| = 2J$. 
Consequently, the perturbative regime contracts quadratically with the size of the system, as it is governed by the condition $\gamma t \ll 1/J^2$, rendering the perturbative approximation rapidly invalid as $J$ increases. 

\begin{figure}[h!]
    \centering
    \includegraphics[width=1.1\linewidth]{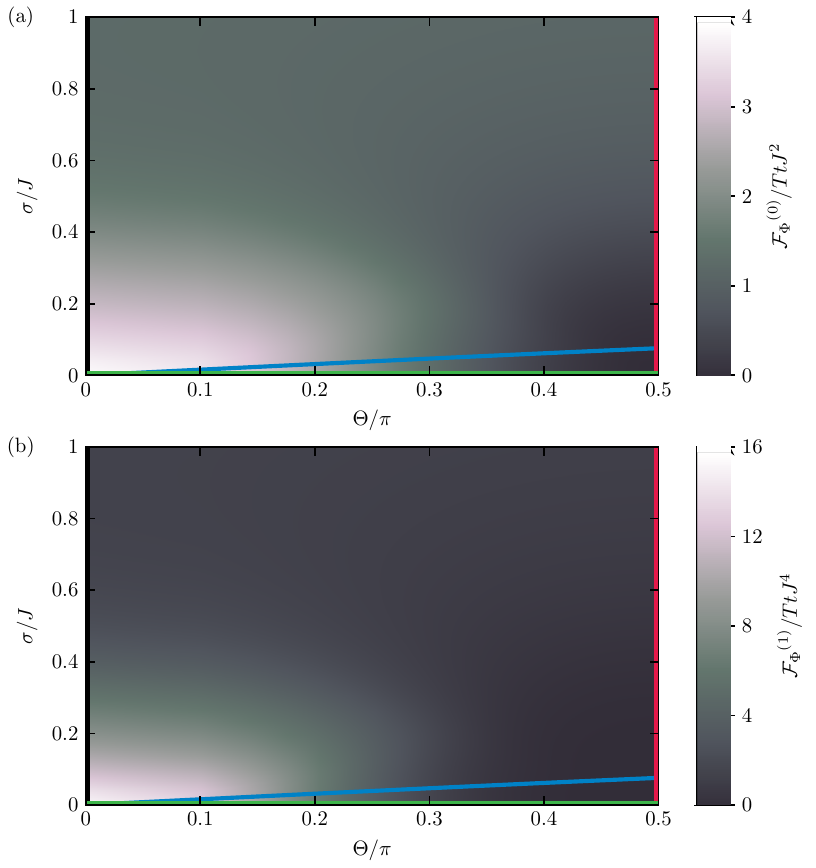}
    \caption{{\bf Sensitivity and degradation indicators for the GSS ansatz.} 
    The sensitivity indicator in~\eqref{eq:F0_correlated} (a) and the degradation indicator in~\eqref{eq:F1_correlated} (b) calculated for the GSS ansatz in~\eqref{eq:probestate_definition}. 
    The improved sensitivity coincides with larger degradation under the noise. 
    Specific families of states are marked with solid lines:  squeezed states (red), SCCSs (blue), superposition of Dicke states (green),  GHZ-like states (black). 
    The data are obtained for $J=20$. 
} 
\label{fig:F0_F1}
\end{figure}

\noindent Although the perturbative expression strictly holds only in the short-time regime, it delivers concrete, easily accessible insight into the precision limit set by the QFI -- in closed analytical form, without the need for a full numerical diagonalization. 
Over longer interrogation timescales, the intuition behind the perturbative formula breaks down, yet the same qualitative story holds: the noise keeps degrading in much the same progressive way as in the sensitivity and robustness measures. 
In Fig.~\ref{fig:F0_F1}, we plot these values for a fixed $J$. 
As expected, the indicators show qualitatively similar behavior in the parameter space, with high sensitivity coinciding with a high degradation indicator. 
Note that the gradient plots have different color scales. 
The values are normalized by the time factor $Tt$ and the spin dimension $J^2$ for $\qfi^{(0)}$ in Fig.~\ref{fig:F0_F1}a and $J^4$ for $\qfi^{(0)}$ in Fig.~\ref{fig:F0_F1}b. 
The different powers of $J$ needed to bring the
two panels onto a comparable scale are themselves the trade-off in visual form. The solid lines in Fig.~\ref{fig:F0_F1} single out the families of states that we study. In the subsequent sections, each of these elements will be examined in greater detail. 

\subsubsection{Spin squeezing}
\label{sec:sss}

Spin-squeezing is the go-to strategy for pushing the estimation precision beyond classical limits.
The typical approach initiates the SCS aligned in the direction $\Theta=\pi/2$ and applies a procedure to reduce the variance in a direction perpendicular to the encoding direction, thus preparing a \emph{spin-squeezed state}~\cite{kitagawa_squeezed_1993, ma_quantum_2011}. \\
For moderate squeezing, the state can be approximated by the GSS state $\ket{\pi/2,\sigma}$ in~\eqref{eq:GSS} where $\sigma$ controls the squeezing. 
Within this ansatz, we recover the SCS for $\mVar = J/2$. 
Squeezing along a direction perpendicular to the encoding corresponds to $\mVar\geq J/2$, that is, an increase of the variance along $\hat z$ and improved sensing of an orthogonal observable. 
On the other hand, $\mVar\leq J/2$ indicates that we squeezed along the $\hat z$ axis thus degrading the precision of the estimation.
Notice that in the extreme case where $\sigma\rightarrow0$, we recover the Dicke state of zero magnetization, that is, $\ketJ{0}$, which is an eigenstate of the encoding operator $\Jz$. 
The variances of the collective spin components orthogonal to the mean spin direction are given by $\Delta^2 \Jz \approx \sigma^2$, which is directly proportional to the $\qfi^{(0)}$ in~\eqref{eq:F0_correlated}, and by $\Delta^2 \Jy \approx\frac{1}{2}
\left[J(J+1)-\sigma^2\right]
\left(1-e^{-1/(2\sigma^2)}\right)$ in the perpendicular direction. 
\\
In the limit of large systems, that is, $J\gg1$, and moderate squeezing, that is, $\sigma/J\ll1$, we calculate the analytical formulas for sensitivity and degradation indicators in~\eqref{eq:F0_correlated} and~\eqref{eq:F1_correlated}, respectively, that is, 
\begin{align}\label{eq:f0_f1_sss}
&\qfi^{(0)}\! \left[ \Psi (\frac{\pi}{2}, \sigma)\right] \approx 4 \nu  t^2\sigma^2, \\
&\qfi^{(1)} \!\left[ \Psi (\frac{\pi}{2}, \sigma)\right] \approx  16  \nu t^2 \sigma^4,
\end{align}
which depend only on the variance $\sigma^2$ and the ratio $\qfi^{(1)}/\qfi^{(0)}\propto \sigma^2$ increases with the variance. 
Despite the impact of squeezing on the QFI degradation, the penalty is not much stronger than the gain. 
Later in Sec.~\ref{sec:optimize_Jy}, we are going to optimize the sensing strategy to obtain the best bound under fixed time of the experiment. 

\begin{figure}[h!]
    \centering
    \includegraphics[width=\columnwidth]{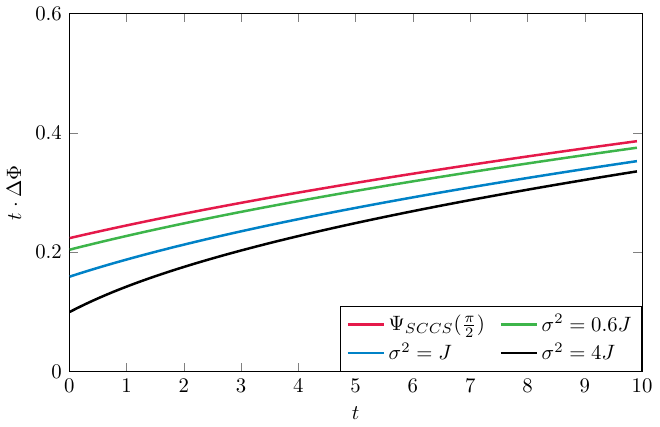}
    \caption{{\bf Spin-squeezed states under spatially correlated dephasing.} 
    { Quantum Cramer-Rao bound as a function of the interrogation time for a squeezed state defined in \eqref{eq:probestate_definition} with $\Theta=\pi/2$ and different variances $\sigma^2$. 
    Parameter encoding occurs in the presence of correlated dephasing described by \eqref{eq:master_eq_correlated}}.
    Data obtained for $\param=1$, spin dimension $J=20$ under dephasing amplitude $\gamma=0.01$.  
    Obtained with exact diagonalization.
    }
    \label{fig:SCS_like}
\end{figure}

\noindent In Fig.~\ref{fig:SCS_like}, we show the full QFI for the spin-squeezed state. The plot illustrates that the short-time dynamics are accurately captured by the sensitivity and degradation indicators. 
The intuition that is based solely on the magnitudes of $\qfi^{(0)}$ and $\qfi^{(1)}$ breaks down at longer times, as is also evident in Fig.~\ref{fig:SCS_like}.
For extended interrogation times, the data presented show that the squeezed state remains more favorable than the SCS state, even in the presence of noise. 

\subsubsection{Superposition of classical states}
\label{sec:scs_cat}

An SCS can be written as a product state of identical spins
\begin{align}\label{eq:SCSmain}
    \ket{\Theta, \varphi} \equiv \bigotimes_{j=1}^N\left(\cos{({\Theta}/{2})}\ket{0}+ \sin{({\Theta}/{2})} e^{i\varphi}\ket{1} \right),
\end{align}
all aligned in the same direction in the Bloch sphere. 
The system consisting of $N$ spins is equivalent to the total spin $J = N/2$. 
The orientation, or mean spin direction, of the state is specified by a polar angle $\Theta$ and an azimuthal angle $\varphi$.
The SCS in~\eqref{eq:SCSmain} is the eigenstate of the spin operator in the same orientation with the maximal eigenvalue. 
SCSs exhibit isotropic quantum fluctuations in the plane orthogonal to the mean spin direction. 
The variance of the collective spin operator in any such perpendicular direction $\vec{n}_\perp$ is given by $\Delta^2 J_{\vec{n}_\perp}=J/2$, where $\hat{J}_{\vec{n}_\perp} = \hat{\vec{J}} \cdot \vec{n}_\perp$ denotes the projection of the collective spin operator along $\vec{n}_\perp$. 
We set $\varphi=0$ in \eqref{eq:SCSmain} to satisfy the spin-flip (mirror) symmetry of the GSS superposition in~\eqref{eq:probestate_definition}. 
\\
In the asymptotic limit, that is, $J\gg1$, the SCS constitutes a particular instance of the GSS
in~\eqref{eq:GSS} with the SCS variance
\begin{align}\label{eq:varSCS}
\mVar_{\rm SCS} = {J \sin^2(\Theta)}/{2} .
\end{align}
Here we study \eqref{eq:probestate_definition} with $\sigma=\sigma_{SCS}$, which we refer to as a spin coherent cat state (SCCS), denoted by $\ket{\Psi_{\rm SCCS}(\Theta)}$. 
\\
The sensitivity, that is, the noiseless QFI in~\eqref{eq:F0_correlated} can be computed analytically, yielding
\begin{align}\label{eq:QFIunitartSCCS}
   &\qfi^{(0)}\left[\Psi_{\rm SCCS}(\Theta)\right] \nonumber \\&=  \nu \frac{t^2J  \left[2J+1 + (2J-1)\cos(2\Theta) + 2\sin^{2J}\Theta\right]}{1+\sin^{2 J}{\Theta}} ,
\end{align}
which increases as $\Theta$ decreases to $0$. 
We can verify that for the product state $\ket{\Psi_{\rm SCCS}(\frac{\pi}{2})}$, the QFI follows the standard quantum limit, that is,  $\qfi^{(0)} = 2J  \nu t^2$, and for the GHZ state $\ket{\Psi_{\rm SCCS}(0)}$ we saturate the Heisenberg limit scaling, that is,  $\qfi^{(0)} = 4  \nu J^2 t^2$.
\\
The robustness measure in~\eqref{eq:F1_correlated} can be computed in a similar way. 
In the limit of large $J$ and for $\Theta\neq\pi/2$, such that $\sin^{2J}\Theta\rightarrow0$, we obtain the approximate formula
\begin{align}\label{eq:qfi1_SCS_cat_largeJMAIN}
\qfi^{(1)}\left[\Psi_{\rm SCCS}(\Theta)\right] \approx&  16 \nu  t^2
J^4  \cos^4\Theta 
\end{align}
and for $\Theta=\pi/2$ we have $\qfi^{(1)}\left[ \Psi_{\rm SCCS}(\pi/2)\right] = 16 J^2 \nu t^2$. The approximation in~\eqref{eq:qfi1_SCS_cat_largeJMAIN} is valid for fixed \(\Theta\neq\pi/2\) in the large-\(J\) limit, but it is not uniform as \(\Theta\) approaches \(\pi/2\). Its accuracy depends on the angle; e.g., for \(\Theta=\pi/4\), \(\sin^{2J}\Theta<10^{-6}\) already when \(J=20\). \\
The expression in \eqref{eq:qfi1_SCS_cat_largeJMAIN} decreases with $\Theta$. For small angles $\Theta$ and $J\gg1$, a direct calculation shows that $\qfi^{(1)}/\qfi^{(0)}\approx 4J^2\cos^2\Theta$ which reveals strong scaling with the spin dimension $J$ and $\Theta$-dependence. 
For a SCS, that is, $\Theta=\pi/2$, the ratio $\qfi^{(1)}/\qfi^{(0)} = 8J$ scales much slower in $J$. 

\begin{figure}[h!]
    \centering
    \includegraphics[width=\columnwidth]{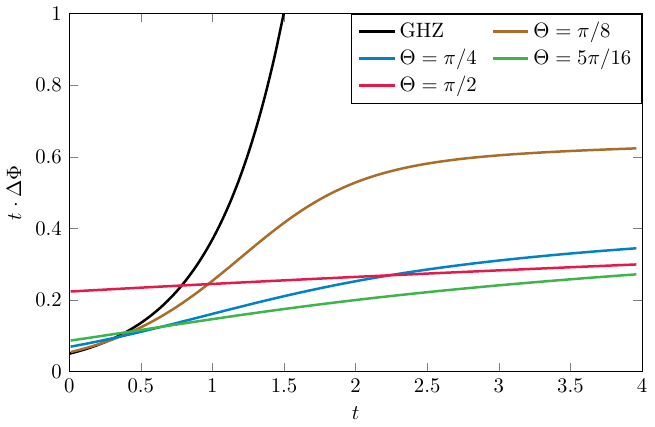}
    \caption{{\bf Spin coherent cat states under spatially correlated dephasing.} 
    Quantum Cram\'er--Rao bound for the SCCSs with different orientation $\Theta$. 
    Data obtained for $\param=1$, spin dimension $J=20$ under dephasing amplitude $\gamma=0.01$.  
    Obtained with exact diagonalization.
    }
    \label{fig:SCS_cat}
\end{figure}

The full QFI dynamics is presented in Fig.~\ref{fig:SCS_cat}. 
For correlated dephasing, it was already shown in~\cite{huang_quantum_2015} that SCCSs can be more robust for metrology than GHZ states. 
Now, with the sensitivity and degradation indicators, we are able to explain the QFI dynamics based on analytical grounds. 
By comparison, the product state exhibits a softer noise correction, making it significantly more robust to dephasing and eventually, for longer interrogation times, reaching better precision than the GHZ state, see Fig.~\ref{fig:SCS_cat}. 
Similar logic holds when we compare two state with different orientation angles $\Theta$. 
Over longer interrogation timescales, the intuition behind the perturbative formula breaks down, yet the same qualitative story holds: the noise keeps degrading in much the same progressive way as in the sensitivity and degradation indicators. 
This illustrates our previous observation: states that offer the strongest quantum enhancement (e.g., GHZ) are also the most fragile under dephasing noise, while states with more modest scaling under noiseless encoding retain their metrological usefulness for longer interrogation times.
Since the SCCS family does not have a single preferred experimentally motivated measurement, we are not going to analyze it. 

\subsubsection{Superposition of Dicke states}
\label{sec:dicke_cat}

In this section, we analyze the limiting behavior of the GSS ansatz in~\eqref{eq:probestate_definition} in the limit $\sigma \to 0$.
There, the GSS superposition simplifies to a coherent superposition of Dicke states
\begin{align}\label{eq:dicke_cat}
\catTh{\Theta, 0} = \normN{\catTh{\Theta, 0}}\left(\ketJ{\mAv} + \ketJ{-\mAv}\right) ,
\end{align}
where $\mAv = \lfloor J\cos\Theta \rfloor$ defines the orientation of the superposition and the normalization is $\normN{\catTh{\pi/2, 0}}=1$ for $\Theta=\pi/2$ and $\normN{\catTh{\Theta, 0}}=1/\sqrt{2}$ otherwise.
\\
The QFI of the Dicke-state superposition in the presence of noise is straightforward to evaluate because the noise preserves the symmetry of the system, and the density matrix remains of rank two.  
A closed-form analytical expression for the QFI reads
\begin{align}\label{eq:qfi_dicke}
 \qfi\left[\Psi(\Theta, 0) \right] = 4 \nu t^2 \mAv^2 e^{-4 \mAv^2 \gamma t} ,
\end{align}
which exhibits exponential suppression due to noise. 
A detailed discussion is provided in App.~\ref{app:Dicke_cat_qfi_fiPar}. 

\begin{figure}[h!]
    \centering
    \includegraphics[width=\columnwidth]{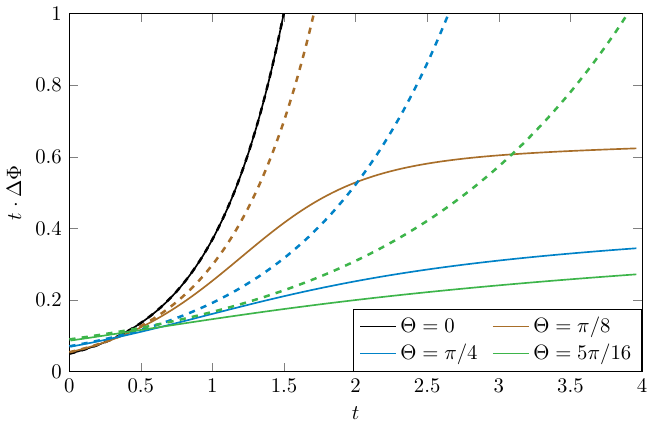}
    \caption{{\bf Comparison between spin coherent cat states (SCCS) and Dicke cat states.}
    SCCS (solid) and Dicke-cat (dashed) with the same orientation $\Theta$. 
    Data obtained for $\param=1$, spin dimension $J=20$ under dephasing amplitude $\gamma=0.01$.  
    Obtained with exact diagonalization.
    }
    \label{fig:Dicke_vs_SCS}
\end{figure}

\noindent An interesting pattern appears when we compare the sensitivity and robustness in~\eqref{eq:F0_correlated} and~\eqref{eq:F1_correlated} for the Dicke-state and SCCS
of Sec.~\ref{sec:scs_cat}. For large $J$, Dicke-state and classical-state superpositions prepared with the same orientation-angle exhibit a remarkably tight correspondence. \\ 
For the Dicke-cat state, the sensitivity and degradation indicators read
\[
\qfi^{(0)} = 4 \nu t^2 J^2\cos^2\Theta, \quad \qfi^{(1)} = 16 \nu t^2 J^4\cos^4\Theta ,
\]
which coincides with the dominant-$J$ term for the SCCS.
\\
The comparison between Dicke superpositions and SCCSs for fixed $J$ is presented in Fig.~\ref{fig:Dicke_vs_SCS}. 
Although Dicke-state superpositions exhibit similar scaling behavior to SCCSs for short timescales,
their metrological usefulness, quantified by the QFI, decays exponentially under noise, a stark
contrast to the more resilient SCCSs. The reason is structural. Both probes are pure, but the
Dicke cat is supported on only two magnetization components, so its entire phase sensitivity rests
on the single coherence $\rho_{k,-k}$, decaying with rate $2 k^2 \gamma$, set by their separation. Once that
coherence is gone, the evolved state is diagonal in the $\Jz$ basis, carries no dependence on $\param$ whatsoever and its QFI is zero. The SCCS
is built from the same two components, but each is now smeared over a range of Dicke states
of width $\sigma_{SCS}$. The coherences between those two components are separated by $\vert m-n \vert \simeq 2J \cos \Theta$, and are destroyed just as fast as those of the Dicke cat. The pairs lying within a single component, however, are separated by only $\vert m-n \vert \sim \sigma_{SCS}$ and survive far longer, so they go on accumulating phase after the others are
gone. Rather than collapsing to zero, the SCCS therefore degrades steadily towards the performance
of a single spin coherent state of variance $\sigma_{SCS}^2.$
 \\
The superposition of Dicke states is further analyzed in Sec.~\ref{sec:optimize_Par}, where we derive the optimal parameter and optimal interrogation time under a restriction of the time resource. 
We demonstrate that the parity measurement saturates the quantum Cram\'er--Rao bound, \textit{i.e.}, it constitutes an optimal measurement, implying the equality of the QFI and the classical FI associated with the parity measurement.

\subsubsection{GHZ and GHZ-like states}
\label{sec:ghz_like}

We extend the notion of GHZ states to a broader family of states given by~\eqref{eq:probestate_definition} with $\Theta = 0$ and arbitrary width $\sigma$. 
We refer to these more general, tunably delocalized states as GHZ-like states. 
This construction reproduces the typical statistics of GHZ-like states~\cite{micheli_many-particle_2003,zhang_fast_2024}, and physically represents an imperfect preparation of the ideal GHZ state.
For small $\sigma$, the GHZ-like states can be viewed as a superposition of two oppositely oriented spin-squeezed states~\cite{kitagawa_squeezed_1993, ma_quantum_2011}. 
\\
Our primary aim is to compare the sensitivity and robustness of the GHZ state with its GHZ-like counterparts. 
In noiseless quantum metrology, GHZ states provide Heisenberg-limited sensitivity, however, even small amounts of noise dramatically degrade this advantage~\cite{huelga_improvement_1997,huang_quantum_2015,smirne_huelga_2016}. 
The quantum Cram\'er--Rao bound derived displays an exponential deterioration of the estimation precision and rapidly deviating from the corresponding noiseless precision, that is, $\qfi \left[ \Psi (0, 0)\right] = 4 \nu t^2 J^2 e^{-4 \gamma t J^2}$, see the expression in~\eqref{eq:qfi_dicke}. 

\begin{figure}[h!]
    \centering
    \includegraphics[width=\columnwidth]{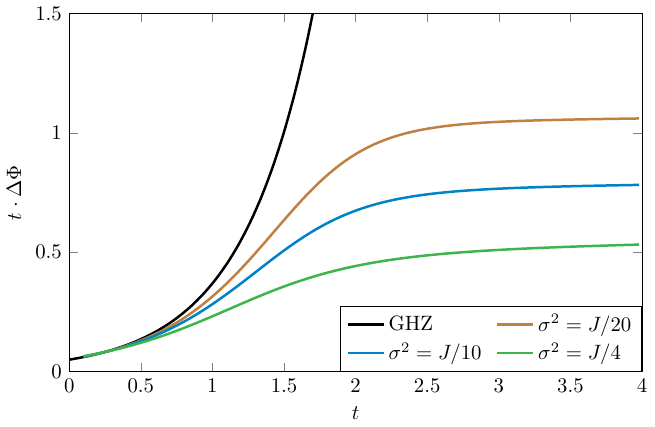}
    \caption{
    { \bf GHZ-like states under spatially correlated dephasing.} 
    {Quantum Cramer-Rao bound as a function of the interrogation time for a GHZ-like state defined in \eqref{eq:probestate_definition} with $\Theta=0$ and different variances $\sigma$. Parameter encoding occurs in the presence of correlated dephasing described by \eqref{eq:master_eq_correlated}}.
    Data obtained for $\param=1$, spin dimension $J=20$ under dephasing amplitude $\gamma=0.01$.  
    Obtained with exact diagonalization.
    }
    \label{fig:GHZ_like}
\end{figure}

\noindent
We characterize the GHZ-like state ansatz by the sensitivity and degradation indicators in~\eqref{eq:F0_correlated} and in~\eqref{eq:F1_correlated}, respectively. 
The sensitivity indicator reads
\[\label{eq:F0_ghzlike}
\qfi^{(0)} \left[ \Psi (0, \sigma)\right] \approx 4 \nu t^2 J^2 \left(1 - 
2\sqrt\frac{2}{\pi}\frac{\sigma}{J}
\right) ,
\]
where in the derivation we used $\sigma/J\ll1$ and $J\gg1$. 
The sensitivity indicator is reduced compared to the GHZ which is intuitive from the geometric picture where the weight of the distribution gets closer to the equator of the Bloch sphere and the variance is reduced. \\
Similarly, we calculate the degradation indicator to be
\[\label{eq:F1_ghzlike}
\qfi^{(1)} \left[ \Psi (0, \sigma)\right] \approx 16 \nu t^2 J^4 \left( 1 -4\sqrt\frac{2}{\pi}\frac{\sigma}{J}\right) , 
\]
where the $\sigma$-dependent correction to $F_\Phi^{(1)}$ is negative, so finite broadening reduces the leading degradation rate.
The ratio of the indicators $\qfi^{(1)}/\qfi^{(0)} \approx 4J^2 \left( 1 -2\sqrt\frac{2}{\pi}\frac{\sigma}{J}\right)$, which suggests $J$-driven domination of the noise impact. 
On the other hand, the $\sigma$-induced correction to $\qfi^{(1)}/\qfi^{(0)}$ is also amplified by the same $J^2$ factor, underscoring how even small deviations from the ideal GHZ state can have a substantial impact.
\\
Figure~\ref{fig:GHZ_like} illustrates the full QFI dynamics for these GHZ-like states. 
Crucially, while the noiseless term also decreases, the noise correction is affected by a larger prefactor $J^4$ for $\qfi^{(1)}$ over $J^2$ for $\qfi^{(0)}$, explaining why GHZ-like states with finite variance can outperform perfectly prepared GHZ states under correlated dephasing.
On the scale of the plot, the correction to the noiseless QFI is almost invisible but the reduction of the noise-induced correction corresponds to a slower QFI degradation, thereby effectively prolonging the metrological usefulness of the state.
\\
The structural similarity to the GHZ state provides a practical motivation for the parity measurement. 
In Sec.~\ref{sec:optimize_Par}, we compare the QFI bound with the classical Cram\'er--Rao bound associated with parity measurements. 
We determine the optimal sensing parameter for the classical bound. 
Within this comparison, we demonstrate the advantages of employing an imperfectly prepared GHZ state for noisy phase estimation.

\section{Optimal sensing strategy}
\label{sec:optimize}

After analyzing the QFI dynamics in the previous section, we now focus on a practical approach to quantum metrology. 
Based on that intuition, we expect a metrological advantage from spin squeezing and, even more likely, from imperfect GHZ state preparation. 
To place this in a practical context, we compare the QFI-based bounds with the measurement-specific bounds for spin-projection and parity measurements that are commonly employed in the experiment.
We optimize the measurement-specific and quantum Cram\'er--Rao bounds under a fixed total experimental time $T$.
For the measurement-specific bounds, we determine the optimal parameter and interrogation time analytically. 
Later, we numerically search for the optimal parameters of the sensing experiment and determine the ultimate precision for each readout. 
For methodological details, see App.~\ref{app:methods}.
\\
In quantum metrology, we infer the unknown parameter $\param$ from the classical data obtained by measuring the probe.
A practically motivated estimation strategy is to record repeated measurements of an observable $\hat O$, with results $\{\mu_1,\mu_2\dots\mu_\nu\}$, and to build the estimate from the {\it sample mean} $M_\nu = \frac{1}{\nu} \sum_{i=1}^\nu \mu_i$ alone.
The sensitivity attainable in this way is given by the error-propagation formula~\cite{pezze2014}
\begin{align}\label{eq:FI_moments}
\cfiEP=\frac{\nu \abs{\partial_\xi \expval{\hat O}_{\xi} }}{\Delta^2 \hat{O}}^2 \!\Bigg|_{\xi=\param} ,
\end{align}
where $\expval{\hat O}_\xi$ and $\Delta^2\hat O$ are evaluated on the evolved state.
The expression in~\eqref{eq:FI_moments} follows from the central limit theorem: for $\nu\to\infty$ the sample mean is Gaussian distributed with mean $\expval{\hat O}_\param$ and variance $\Delta^2\hat O/\nu$, so that error propagation yields the estimator variance $\Delta^2\tilde\param = \Delta^2\hat O/(\nu\abs{\partial_\param \expval{\hat O}}^2)$.
We emphasize that $\cfiEP$ is the sensitivity associated with the estimated mean and variance of $\hat O$, and not, in general, the classical FI of the full distribution of outcomes of $\hat O$.
For a multi-outcome observable such as $\Jy$, the sample mean discards the information carried by the higher moments of the outcome distribution, so that~\eqref{eq:FI_moments} is a lower bound, $\cfiEP\leq\cfi$, and the associated Cram\'er--Rao bound is correspondingly an upper bound on the achievable precision.
The two quantities coincide whenever $\hat O$ has only two outcomes, since the mean then determines the outcome distribution completely.
This is precisely the case for the parity measurement of Sec.~\ref{sec:Par}, for which~\eqref{eq:FI_moments} is the exact classical FI $\cfi$ and we use that notation throughout.
For the fixed duration of the full experiment, the number of measurements that we collect is given by the fraction $\nu=T/t$, where $t$ is the interrogation time and $T$ is the total time resource.
The Cram\'er--Rao bound for the estimation precision is given by the expression in~\eqref{eq:cramer_rao}.
\\
In this part, we inject physically motivated sensing strategies to benchmark against the optimal sensitivity, revealing concrete, practical insights into the families of states we study. 
We consider spin projection and parity measurements, common strategies known to saturate the noiseless quantum Cram\'er–Rao bound for the spin coherent state and the GHZ state, respectively. 

\subsection{Spin projection measurement}
\label{sec:Jy}

The spin projection measurement is one of the most commonly used approaches in the experimental scenario. 
For the classical spin state oriented orthogonally to the encoding, \textit{i.e.}, in our case $\ket{+}^{\otimes 2J}$ where $\ket{+}$ is the eigenstate of the matrix $\sigma^x$, the operator $\Jy$ is an optimal estimator, see App.~\ref{app:scs_optimal_is_Jy}. 
Spin-projection measurements are widely utilized for phase estimation with spin-squeezed states, however, associated estimators generally do not reach the quantum Cram\'er--Rao bound that can be achieved with an optimal measurement strategy. 
In the presence of noise, spin-projection measurements are no longer optimal even for a spin coherent state, yet they continue to be employed as a practically feasible measurement protocol.
\\
The error-propagation sensitivity in~\eqref{eq:FI_moments} associated with a measurement of the operator $\Jy$ can be evaluated analytically using the full expression
\begin{align}
\label{eq:FI_Jy}
&\cfiEP = \! \frac{2 T t e^{-\gamma t} \cos^2\param t \expval{\Jx}^{\!2}}{
 f_-\expval{\Jx^2}\!+\! f_+\Delta^2\Jy \!-\! 2 e^{-\gamma t}\sin^2\param t  \expval{\Jx}^{\!2}},
\end{align}
where $f_\pm=1\pm e^{-2\gamma t}\cos2\param t$ and the observables are calculated for the input probe state. 
The sensitivity $\cfiEP$ is maximized for the optimal parameter $\paramOpt=n\pi/t$, where $n$ is an integer.
For the optimal parameter, it reads
\begin{align}\label{eq:fi_paramOpt_Jy}
\cfiEP\overset{\paramOpt}{=}& \frac{T t \expval{\Jx}^2}{\cosh(\gamma t) \Delta^2\Jy^2 +\sinh(\gamma t) \expval{\Jx^2}}  .
\end{align}
The moment-based sensitivity can be further optimized with respect to the interrogation time. 
The optimal time depends on the variance of the initial state and reads 
\[\label{eq:topt_Jy_main}
\tOpt \overset{\paramOpt}{\approx} \frac{1}{\gamma}\left(\frac{3\Delta^2\Jy}{\expval{\Jx^2}}\right)^{\!1/3},
\]
which is derived for moderate squeezing, that is, $\sigma\ll J$, and large spin dimension, that is, $J\gg 1$. 
For a more extended comment on the derivation, see App.~\ref{app:Jy}. 
\\
Finally, the optimized sensing strategy can be used to calculate the ultimate precision for the spin-projection readout
\begin{align}\label{eq:dPhi_paramOpt_topt_Jy}
\Delta^2\paramOpt\overset{\tOpt}{\gtrsim}& \frac{\gamma}{T}\left[1+\frac{1}{2}\left(\frac{3\Delta^2\Jy}{\expval{\Jx^2}}\right)^{\!2/3}\right] .
\end{align}
The formula in~\eqref{eq:dPhi_paramOpt_topt_Jy} exposes a fundamental probe-independent sensitivity limit: $\Delta^2\paramOpt\geq \gamma / T$ with a correction term that depends on squeezing along the $\Jy$ direction and system size.
The lowest achievable bound depends on the noise amplitude and cannot be beaten even if the squeezing and system size increase. 
In fact, these parameters enhance the precision, however, the improvement is only by boosting the subleading term. 
\\
For the classical state, where $\Delta^2\Jy=J/2$, we can verify that $\Delta^2 \paramOpt\overset{\tOpt}{\gtrsim}\frac{\gamma}{T}\left[1+ \frac{1}{2}\left(\frac{3}{2}\right)^{\!2/3}J^{\!-2/3}\right]$. 
For the optimal state in two axis twisting~\cite{kitagawa_squeezed_1993}, where $\Delta^2\Jy\sim \frac{1}{2}$, the precision bound achieves
$\Delta^2 \paramOpt\overset{\tOpt}{\gtrsim}\frac{\gamma}{T}\left[1+ \frac{1}{2}\left(\frac{3}{2}\right)^{\!2/3}J^{\!-4/3}\right]$. 
In the next section, we will extend this consideration to an arbitrary squeezed state and compare the optimal error-propagation and quantum Cram\'er--Rao bounds to each other.

\subsubsection{Spin squeezing}
\label{sec:optimize_Jy}

For large system sizes and moderate squeezing, that is, $J\gg1$ and $\sigma/J\ll1$, we can calculate the error-propagation Cram\'er--Rao bound in terms of the GSS variance using the general expression in~\eqref{eq:dPhi_paramOpt_topt_Jy} which is extremized for the parameter $\paramOpt=\frac{n\pi}{t}$ and the interrogation time $\tOpt$ in~\eqref{eq:topt_Jy_main}.
The bound reads
\begin{align}\label{eq:d2Phi_paramOpt_tOpt_Jy_sssMAIN}
\Delta^2\paramOpt\overset{\tOpt}{\gtrsim}& \frac{\gamma}{T}\left[1+\frac{1}{2}\left(\frac{3}{4}\right)^{2/3}\!\!\sigma^{-4/3}\right],
\end{align}
where the expression indicates an improvement in the phase estimation with the squeezing, that is, an increase in the variance $\sigma$. 
Notice that for the GSS ansatz, precision is still lower-bounded by $\gamma/T$, but the subleading term is entirely controlled by $\sigma$. 
For more extended comment, see App.~\ref{app:sss_Jy}. 

\begin{figure}[h!]
    \centering
    \includegraphics[width=\columnwidth]{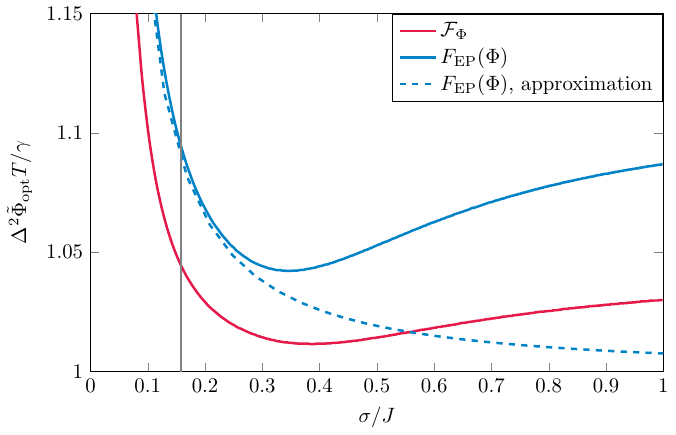}
    \caption{
    {\bf Optimized Cram\'er--Rao bounds for squeezed state. }
    Comparison of the optimized error-propagation and quantum Cram\'er--Rao bounds for the squeezed state modeled by the GSS in~\eqref{eq:probestate_definition} with $\Theta=\pi/2$ as a function of the normalized variance $\sigma/J$.
    The bound is expressed for a fixed total experiment time $T$ and dephasing $\gamma$.
    The quantum Cram\'er--Rao bound (red) achieves higher precision than the moment-based estimation using the spin projection measurement (blue).
    The approximation in~\eqref{eq:d2Phi_paramOpt_tOpt_Jy_sssMAIN} (dashed blue) is a good approximation about the SCS state (marked with gray vertical line). 
    Data obtained for numerically optimized phase parameter and interrogation time at $J=20$ under correlated dephasing $\gamma=0.01$. 
    }
    \label{fig:FI_SCS_like_main}
\end{figure}

\noindent
Figure~\ref{fig:FI_SCS_like_main} shows the numerically calculated and optimized error-propagation and quantum Cram\'er--Rao bounds.
The bounds are plotted as a function of the GSS superposition variance for a fixed system size.
In the presence of noise, the $\Jy$ operator is not an optimal observable even for a SCS, explaining the mismatch between the quantum and the error-propagation bounds for the classical state in Fig.~\ref{fig:FI_SCS_like_main}.
Initially, increasing squeezing improves precision, but beyond an optimal $\sigma$, it reduces estimation accuracy for both bounds. 
The over-squeezing leads to a reduction in the attainable precision in both cases.
The expression in~\eqref{eq:d2Phi_paramOpt_tOpt_Jy_sssMAIN} accurately captures the improvement in the moment-based estimation precision around the SCS.
The expression does not capture the turnover in $\Delta^2\paramOpt$. 

\subsection{Parity measurement}
\label{sec:Par}

Parity measurements are directly linked to experimentally accessible observables across diverse quantum platforms, including trapped ions, ultracold atomic ensembles, and nitrogen-vacancy centers in diamond.
For GHZ states and other superpositions of Dicke states, it is well-known that the parity measurement $\hat{O} = \hat\sigma_x^{\otimes N}$ is the optimal choice for unitary encoding, that is $\gamma=0$ in~\eqref{eq:master_eq_correlated}. 
In this setting, the parity readout saturates the Cramér–Rao bound and reaches the Heisenberg limit, achieving maximal precision scaling.
In the presence of dephasing, the parity measurement remains the optimal estimator for all superpositions of Dicke states but is not optimal for GHZ-like states. 
However, owing to their similarity to the ideal GHZ state, parity measurements are also investigated as a practically motivated choice.
\\
The states within a symmetric subspace and spin-flip (mirror) symmetry, that is, $\ket{\Psi}=\sum_m c_m\ketJ{m}$ with $c_m=c_{-m}$, have even parity, that is, $\Par\ket{\psi} = \ket{\psi}$ for the parity operator $\Par$. 
There, the Fisher information for our model in~\eqref{eq:master_eq_correlated} reads
\[\label{eq:cfi_Par_XYmirror_main}
\cfi=\nu\frac{\abs{\partial_\xi \expval{\hat \Pi(t)}_{\xi} }}{1 -  \expval{\hat \Pi(t)}^2}^2  \eval_{\xi=\param} ,
\]
where $\nu=T/t$ and
\[
\hat \Pi(t) = \sum_{m=-J}^J e^{-2\gamma t m^2 + 2\im\param t m}\ketJ{m}\braJ{m} ,
\]
encodes the $\param$ and noise impact $\gamma$. 
For a more detailed discussion, see App.~\ref{app:Par}. 

\noindent Let's dive into a revealing example that exposes the limitations of this sensing strategy by considering the GHZ states. 
Again, we are going to analyze the sensing precision under the restriction to a total sensing time $T$.  
The parity measurement saturates the quantum Cram\'er--Rao bound for GHZ states in the presence of correlated dephasing noise. 
{In the present case, \textit{i.e.} under spatially correlated noise}, the Cram\'er--Rao bound reads
\begin{align}
\!\!\Delta^2 \param \geq \frac{e^{4 \gamma  J^2 t} \csc ^2(2 \param  J t) \left[1\!-\!e^{-4 \gamma  J^2 t} \cos ^2(2 \param  J t)\right]}{4 J^2 t T},
\end{align}
which is minimized for the optimal parameter $\paramOpt=\frac{\pi}{4Jt}$ where the precision reaches 
\begin{align}
\Delta^2 \paramOpt {\geq} \frac{e^{4 \gamma  J^2 t}}{4 J^2 t T}.    
\end{align} 
The strategy can be further optimized by choosing the optimal interrogation time 
\[
\tOpt \overset{\paramOpt}{=} \frac{1}{4\gamma J^2}.\]
With this, we arrive at the ultimate precision bound
\begin{align} \label{eq:GHZopt_CRB_main}
    \Delta^2 \paramOpt \overset{\tOpt}{\geq} \frac{e \gamma}{T} ,
\end{align}
which is independent of the system size and sets the classical and quantum Cram\'er--Rao bounds at the same time. 
For any other superposition of Dicke states, the precision is also bounded by the expression in~\eqref{eq:GHZopt_CRB_main}, see App.~\ref{app:Dicke_cat_qfi_fiPar}. 
\\
Interestingly, the precision bound in~\eqref{eq:GHZopt_CRB_main} is larger than the bounds calculated for spin-projection measurement in~\eqref{eq:dPhi_paramOpt_topt_Jy} by a factor of the Euler number $e\approx2.72$. 
This implies that, in the noisy scenario, the GHZ states achieve worse precision than the classical states. 
\\
Beyond the GHZ ansatz, the analysis becomes more involved. 
To see the parameter and time dependence of the FI in~\eqref{eq:cfi_Par_XYmirror_main}, we expand the expression to obtain
\begin{align}\label{eq:fi_Par}
    \cfi = 4 T t \frac{\left|\sum_m|c_m|^2 m e^{-2\gamma t m^2}\sin(2\param t m)\right|^2} {1 - \left(\sum_m|c_m|^2 e^{-2\gamma t m^2}\cos(2\param t m)\right)^2} ,
\end{align}
where $c_m$ is normalized distribution over the coefficients. 
From the expression in~\eqref{eq:fi_Par}, the phase accumulation rate clearly depends on the component $\ketJ{m}$ and the same holds for the impact of dephasing. 
In contrast to the case of spin-projection measurements, the Fisher information associated with the parity measurement in~\eqref{eq:cfi_Par_XYmirror_main} in general does not admit a straightforward factorization, which can be used to determine the optimal parameter and interrogation time. 
Nevertheless, for GHZ-like states sufficiently close to the GHZ state, we can simplify the analytical expressions for the bound.

\subsubsection{GHZ-like states}
\label{sec:optimize_Par}

We analyze the FI expression in the limit of $\sigma\ll J$ and $J\gg 1$ where the continuum of the sum can be applied. 
We outline the full derivation in the App.~\ref{app:ghz_like_par}. 

Analyzing the extremum of the Fisher information we find that the optimal parameter is approximately
\[
\paramOpt \approx \frac{\pi}{4 J t}\left(1 + \sqrt\frac{2}{\pi}\frac{\sigma}{J}\right), 
\]
where the optimal parameter admits a linear shift in terms of $\sigma/J$.
Assuming the optimal parameter, we repeat the procedure to find the optimal interrogation time
\[
\tOpt \overset{\paramOpt}{\approx} \frac{1}{4 J^2\gamma}\left(1 + 2\sqrt\frac{2}{\pi}\frac{\sigma}{J}\right), 
\]
which also gets a linear correction to the GHZ-optimal point. 

\noindent Finally, we can write the optimal parity FI for the GHZ-like states to be
\begin{align}
\cfi \overset{\paramOpt, \tOpt}{\approx}& \frac{T}{e\gamma}\left(1- \frac{(\pi-2)(8+\pi^2)}{4\pi} \left(\frac{\sigma}{J}\right)^2\right) \\\approx& \frac{T}{e\gamma}\left(1- 1.6234\, \left(\frac{\sigma}{J}\right)^2\right) , 
\end{align}
which gets a quadratic correction to the optimal FI for the GHZ states that decreases the estimation precision as the broadening increases. 
The corresponding Cram\'er--Rao bound reads 
\begin{align}\label{eq:d2Phi_paramOpt_tOpt_Par_ghz_likeMAIN}
\Delta^2\paramOpt \overset{\tOpt}{\approx}& \frac{e\gamma}{T}\left(1+ \frac{(\pi-2)(8+\pi^2)}{4\pi} \epsilon^2\right) \\\approx& \frac{e\gamma}{T}\left(1+ 1.6234\, \epsilon^2\right) .
\end{align}
Relative to conventional GHZ states, the optimal measurement precision for the GHZ-like family receives an additional correction that depends on the state’s variance. 
More precisely, it is displaced by a contribution proportional to the variance of the GHZ-like state, where the change in the GHZ variance scales as $\mathcal{O}\!\left(\frac{\sigma}{J}\right)^2$ and becomes progressively less significant as $J$ grows. 
This shows that, in the large‑$J$ limit, every GHZ‑like state attains GHZ-level sensitivity. 

\begin{figure}[h!]
    \centering
    \includegraphics[width=\columnwidth]{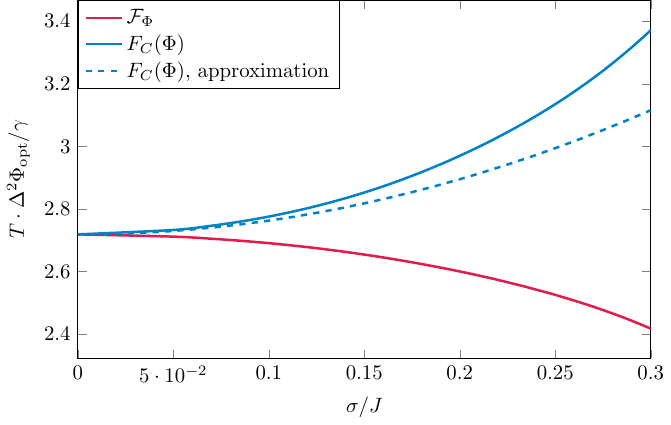}
    \caption{
    {\bf Optimized Cram\'er--Rao bounds for  GHZ-like state. }
    Comparison of the optimized classical and quantum Cram\'er--Rao bounds for the GHZ-like states modeled by the GSS in~\eqref{eq:probestate_definition} with $\Theta=0$ as a function of the normalized variance $\sigma/J$. 
    The bound is expressed for a fixed total experiment time $T$ and dephasing $\gamma$. 
    The quantum Cram\'er--Rao bound (red) achieves higher precision than the classical estimation using parity measurement (blue). 
    The approximation in~\eqref{eq:d2Phi_paramOpt_tOpt_Par_ghz_likeMAIN} (dashed blue) is a good approximation for small broadening $\sigma/J$. 
    Data obtained for numerically optimized phase parameter and interrogation time at $J=20$ under correlated dephasing $\gamma=0.01$. 
    }
    \label{fig:FI_GHZ_like_main}
\end{figure}

\noindent In Fig.~\ref{fig:FI_GHZ_like_main}, we compare the optimized precision bounds for classical and quantum estimation when the total experiment time $T$ is constrained. 
The degradation of the estimation precision for the parity measurement matches the analytical prediction above; however, the expression in~\eqref{eq:d2Phi_paramOpt_tOpt_Par_ghz_likeMAIN} is valid only in the regime $\sigma/J \ll 1$. 
Remarkably, for optimal measurement, we find improved precision bounds as we diverge from the GHZ state. 
In other words, the imperfect GHZ state actually serves as a superior probe for noisy phase estimation in the presence of dephasing governed by~\eqref{eq:master_eq_correlated}.
{The two observations should not be confused: the QFI advantage of GHZ-like states over the ideal GHZ state is intrinsic to the probe geometry, but it is \emph{not} captured by the parity readout, whose performance degrades with $\sigma/J$ because parity is tailored to the ideal GHZ ansatz at $\sigma=0$. Extracting the QFI advantage therefore requires adapting the measurement strategy to the finite broadening, rather than retaining the parity protocol optimal for the ideal GHZ state.}

\section{Conclusions}
In this paper, we investigated phase estimation in a noisy setting where the noise and the encoding direction are perfectly aligned. Within this framework, we introduced analytical sensitivity and degradation indicators that quantify how the encoding process and the noise affect a given probe state. In the perturbative regime, where the noise-induced evolution remains small over the interrogation time, these measures provide a systematic characterization of the QFI without requiring its full evaluation, which is analytically difficult and computationally demanding in generic many-body settings.
We analyzed the QFI dynamics and its associated indicators for widely studied spin-state families, with a particular focus on several classes of symmetric states, including spin-squeezed states, superpositions of spin coherent states, superpositions of Dicke states, and GHZ-like states.
\\
Beyond comparing specific probe families, our perturbative framework makes explicit how, in the commuting correlated-dephasing setting considered here, the same collective-spin moments that enhance noiseless sensitivity also determine the leading noise-induced loss of QFI. In this sense, our results provide a concrete, state-structural realization of the limitations that arise when signal encoding and noise are generated by the same operator: the degrees of freedom carrying metrological advantage are also those most strongly affected by dephasing. This clarifies, at the level of symmetric large-spin probes, why passive probe design alone cannot eliminate the sensitivity/robustness trade-off in this setting, and identifies state geometry as the key factor governing metrological performance under commuting noise.\\
The framework developed here also provides a systematic way to relate metrological performance to experimentally accessible observables, enabling the identification of suitable probe states without relying on full optimization of the QFI. While the present analysis is restricted to the perturbative regime and to commuting noise models, it suggests a route toward extending such moment-based approaches to more general noise scenarios.
\\
For those cases in which a practically motivated measurement strategy can be formulated, we compared the quantum Cram\'er--Rao bound with the precision attainable using spin-projection and parity measurements. 
The corresponding measurement strategies are not generally optimal, however, they are motivated by experimental accessibility. 
For squeezed states, both {spin-projection-measurement} and quantum bounds improve with increasing squeezing {until the state becomes oversqueezed and the precision degrades}. 
For GHZ-like states, we found that parity-based protocols degrade rapidly away from the ideal GHZ limit, while the QFI can be enhanced by finite broadening, indicating that imperfect GHZ preparation can improve robustness. 
\\
Several concrete extensions follow naturally from this framework. First, the observation that the parity readout is tailored to the ideal GHZ ansatz at $\sigma=0$ motivates a systematic search for $\sigma$-adapted measurement strategies that saturate the QFI of GHZ-like probes, thereby converting the intrinsic QFI advantage of finite broadening into experimentally accessible precision gains. Second, the present analysis treats $\Phi$ as the sole unknown, while in realistic settings the dephasing strength $\gamma$ is itself imperfectly known; extending the moment-based framework to the joint estimation of phase and noise parameters~\cite{kobus_asymptotically_2025} would clarify how the sensitivity-robustness trade-off identified here interacts with the simultaneous estimation of nuisance parameters. Because the moment-based framework relies only on low-order collective-spin expectation values, it offers a natural meeting point between theoretical probe design and experimental certification in atomic-ensemble and trapped-ion platforms, where these moments are directly accessible.

\section{Acknowledgements}
\label{sec:acknowledgements}

G.W. thanks Lennart Bosch, 
Philipp Treutlein, 
Tobias Haas, 
Rafał Demkowicz-Dobrzański 
for fruitful discussions that improved the quality of the paper. 
This work was supported by the ERC Synergy grant HyperQ (Grant No. 856432), the EU project C-QuENS (grant no. 101135359) and QuMicro (Grant No. 101046911). G.~W. acknowledges the Alexander von Humboldt Foundation for support under the Humboldt Research Fellowship. G.~W. acknowledges support by Financial Support Programmes for Early Career Researchers, Graduate and Professional Training Center, Ulm University. 

\appendix

\section{Perturbative QFI under collective dephasing}
\label{app:qfi_perturbative_dephasing}

In this appendix we derive the short-time expansion of the QFI for the collective dephasing model considered in the main text. 
We focus on the case in which both the encoding Hamiltonian and the Lindblad operator are proportional to the same collective spin operator,
\[
G=\Jz .
\]
The encoded noisy state can be written as
\[
\rho_\Phi(\epsilon)
=
e^{-i\Phi t G}\rho_\epsilon e^{i\Phi t G},
\qquad
\epsilon=\gamma t,
\]
where
\[
\rho_\epsilon
=
e^{\epsilon\mathcal D_G}\rho_0,
\qquad
\rho_0=\ket{\psi}\bra{\psi},
\]
with \(\braket{\psi}{\psi}=1\), and
\[
\mathcal D_G[\rho]
=
G\rho G-\frac12\{G^2,\rho\}.
\]
Equivalently, in the eigenbasis of \(G\), with \(G\ket{m}=m\ket{m}\) and \(\ket{\psi}=\sum_m c_m\ket{m}\), the noisy state has matrix elements
\[
(\rho_\epsilon)_{mn}
=
c_m c_n^*
\exp\left[-\frac{\epsilon}{2}(m-n)^2\right].
\]
Since the parameter \(\Phi\) is encoded unitarily, the QFI can be evaluated from the spectral decomposition
\[
\rho_\epsilon=\sum_\alpha \lambda_\alpha \ket{e_\alpha}\bra{e_\alpha}
\]
as
\begin{align}
F_\Phi(\rho_\epsilon)
=
2\nu t^2
\sum_{\alpha,\beta}
\frac{(\lambda_\alpha-\lambda_\beta)^2}
{\lambda_\alpha+\lambda_\beta}
\left|
\bra{e_\alpha}G\ket{e_\beta}
\right|^2 ,
\label{eq:spectral_qfi_unitary}
\end{align}
where terms with \(\lambda_\alpha+\lambda_\beta=0\) are omitted. Here \(F_\Phi\) is the total QFI for \(\nu\) independent repetitions. \\
We now derive the eigenvectors $\{ \ket{e_\alpha}\}$ and eigenvalues $\lambda_\alpha$ to first order in perturbation theory. Let
\begin{align}
\rho_\epsilon &=\rho_0+\epsilon K+O(\epsilon^2), \\ 
K &= \mathcal D_G[\rho_0]=G\rho_0 G-\frac12\{G^2,\rho_0\} \label{eqapendix:K}.
\end{align}
\paragraph{Non-degenerate subspace.} The unperturbed state \(\rho_0\) has one nonzero eigenvalue equal to \(1\), with eigenvector \(\ket{\psi}\), while all remaining eigenvalues are zero. 
Since the eigenvalue \(1\) is non-degenerate, its first-order correction can be obtained from ordinary perturbation theory.\\
We write the perturbed eigenvalue connected to the unperturbed pair \((1,\ket{\psi})\) as
\[
\lambda_0(\epsilon)
=
1+\epsilon\lambda_0^{(1)}+O(\epsilon^2),
\]
and
\[
\ket{e_0(\epsilon)}
=
\ket{\psi}
+
\epsilon\ket{\eta}
+
O(\epsilon^2).
\]
The vector \(\ket{\eta}\) denotes the first-order correction to the eigenvector. 
We choose $\ket{\eta}$ to be orthogonal, \textit{i.e.} $\braket{\psi}{\eta}=0$, which can always be imposed at this order. The eigenvalue equation is
\[
\rho_\epsilon\ket{e_0(\epsilon)}
=
\lambda_0(\epsilon)\ket{e_0(\epsilon)}.
\]
Substituting the perturbative expansions gives
\[
\left(\rho_0+\epsilon K\right)
\left(\ket{\psi}+\epsilon\ket{\eta}\right)
=
\left(1+\epsilon\lambda_0^{(1)}\right)
\left(\ket{\psi}+\epsilon\ket{\eta}\right)
+
O(\epsilon^2).
\]
Keeping terms up to first order in \(\epsilon\), the left-hand side is
\[
\rho_0\ket{\psi}
+
\epsilon\rho_0\ket{\eta}
+
\epsilon K\ket{\psi},
\]
while the right-hand side is
\[
\ket{\psi}
+
\epsilon\ket{\eta}
+
\epsilon\lambda_0^{(1)}\ket{\psi}.
\]
Using $\rho_0\ket{\psi}=\ket{\psi}$, the zeroth-order terms cancel, and the first-order equation becomes
\[
\rho_0\ket{\eta}+K\ket{\psi}
=
\ket{\eta}
+
\lambda_0^{(1)}\ket{\psi}.
\]
Multiplying from the left by \(\bra{\psi}\) we obtain
\[
\lambda_0^{(1)}
=
\bra{\psi}K\ket{\psi}.
\]
Therefore
\[
\lambda_0(\epsilon)
=
1+\epsilon\bra{\psi}K\ket{\psi}
+
O(\epsilon^2).
\]
A direct evaluation gives
\[
\lambda_0^{(1)}
=
\langle G\rangle^2 - \langle G^2\rangle
=
-(\Delta G)^2,
\]
where $(\Delta G)^2=\langle G^2\rangle-\langle G\rangle^2$ and all
expectation values are taken in $\ket{\psi}$.  Hence
\begin{equation}
\lambda_0(\epsilon)
=
1-\epsilon\,(\Delta G)^2+O(\epsilon^2).
\label{eq:lambda0}
\end{equation}

Since $\braket{\psi}{\eta}=0$ implies $\rho_0\ket{\eta}=0$,
the first-order equation reduces to
\[
K\ket{\psi}=\ket{\eta}+\lambda_0^{(1)}\ket{\psi}.
\]
Defining $\delta G\equiv G-\langle G\rangle$, one obtains after
simplification
\begin{equation}
\ket{\eta}
=
-\frac{1}{2}\bigl[(\delta G)^2-(\Delta G)^2\bigr]\ket{\psi},
\label{eq:eta}
\end{equation}
which manifestly satisfies $\braket{\psi}{\eta}=0$.

\paragraph{Degenerate zero-eigenvalue subspace.}
The subspace orthogonal to $\ket{\psi}$ is
degenerate with unperturbed eigenvalue~$0$. We assume \((\Delta G)^2>0\) in what follows. If \((\Delta G)^2=0\), then \(\ket{\psi}\) is a \(G\)-eigenstate and \(F_\Phi=0\) exactly. Letting $P_\perp=I-\ket{\psi}\bra{\psi}$, for any
$\ket{f_j},\ket{f_k}\perp\ket{\psi}$ the anticommutator terms in $K$
vanish and
\[
\bra{f_j}K\ket{f_k}
=
\bra{f_j}G\ket{\psi}\bra{\psi}G\ket{f_k},
\]
so $P_\perp K P_\perp=(P_\perp G\ket{\psi})(\bra{\psi}G P_\perp)$ is
rank one.  Writing
$P_\perp G\ket{\psi}=\delta G\,\ket{\psi}$, its single nonzero
eigenvalue is
\[
\lambda_1^{(1)}
=
\|\delta G\,\ket{\psi}\|^2
=
(\Delta G)^2 ,
\]
with normalised eigenvector
\begin{equation}
\ket{\phi}
=
\frac{\delta G\,\ket{\psi}}{\Delta G}\,.
\label{eq:phi}
\end{equation}
All remaining vectors \(\ket{f_j}\) (\(j\ge2\)) span the residual null subspace orthogonal to both \(\ket{\psi}\) and \(\ket{\phi}\), and satisfy \(\lambda_j^{(1)}=0\). To summarise, the first-order spectrum is
\begin{align}
\lambda_0 &= 1-\epsilon\,(\Delta G)^2+O(\epsilon^2),
&\ket{e_0}&=\ket{\psi}+\epsilon\ket{\eta}+O(\epsilon^2),
\nonumber\\
\lambda_1 &= \epsilon\,(\Delta G)^2+O(\epsilon^2),
&\ket{e_1}&=\ket{\phi}+O(\epsilon), \nonumber
\label{eq:perturbed_spectrum}\\
\lambda_{j\ge2} &= O(\epsilon^2),
&\ket{e_j}&=\ket{f_j}+O(\epsilon),
\nonumber
\end{align}
with $\lambda_0+\lambda_1=1+O(\epsilon^2)$, consistent with trace
preservation.

\paragraph{Substitution into the QFI.}
Using $\lambda_\alpha+\lambda_\beta>0$ and the symmetry of the
summand, ~\eqref{eq:spectral_qfi_unitary} is equivalent to
\begin{equation}
F_\Phi
=
4 \nu t^2
\sum_{\alpha<\beta}^{\lambda_\alpha+\lambda_\beta>0}
\frac{(\lambda_\alpha-\lambda_\beta)^2}
     {\lambda_\alpha+\lambda_\beta}\,
\bigl|\!\bra{e_\alpha}G\ket{e_\beta}\!\bigr|^2.
\label{eq:qfi_half}
\end{equation}
We now classify the pairs $(\alpha,\beta)$ that contribute at first
order in~$\epsilon$.  We denote the central moments
$\mu_k\equiv\langle(\delta G)^k\rangle$ throughout.

\medskip\noindent
\emph{Pair $(\alpha,\beta)=(0,1)$}:
Since $\lambda_0+\lambda_1=1+O(\epsilon^2)$ and
$\lambda_0-\lambda_1=1-2\epsilon\,(\Delta G)^2+O(\epsilon^2)$,
\begin{equation}
\frac{(\lambda_0-\lambda_1)^2}{\lambda_0+\lambda_1}
=
1-4\epsilon\,(\Delta G)^2+O(\epsilon^2).
\label{eq:weight01}
\end{equation}
To obtain $|\bra{e_0}G\ket{e_1}|^2$ to first order without computing
the correction to $\ket{e_1}$, we note that
$\bra{\psi}G\ket{f_j}=0$ for all $j\ge2$ (since
$P_\perp G\ket{\psi}\propto\ket{\phi}\perp\ket{f_j}$), which implies
$|\bra{e_0}G\ket{e_j}|^2=O(\epsilon^2)$ for $j\ge2$.
We notice that by completeness $\bra{e_0}G^2\ket{e_0} = \sum_{j} \bigl|\!\bra{e_0}G\ket{e_j}\!\bigr|^2$. 
Hence, 
\begin{equation}
\bigl|\!\bra{e_0}G\ket{e_1}\!\bigr|^2
=
\bra{e_0}G^2\ket{e_0}
-\bigl|\!\bra{e_0}G\ket{e_0}\!\bigr|^2
+O(\epsilon^2).
\label{eq:completeness_trick}
\end{equation}
Using $\ket{e_0}=\ket{\psi}+\epsilon\ket{\eta}$
and~\eqref{eq:eta}, a direct calculation gives
$\bra{\psi}G\ket{\eta}=-\mu_3/2$ and
\begin{equation}
\bra{e_0}G^2\ket{e_0}
-\bigl|\!\bra{e_0}G\ket{e_0}\!\bigr|^2
=
(\Delta G)^2
+\epsilon\bigl[(\Delta G)^4-\mu_4\bigr]
+O(\epsilon^2).
\label{eq:G01_sq}
\end{equation}
Multiplying \eqref{eq:weight01} and~\eqref{eq:G01_sq}, the $(0,1)$
pair contributes
\begin{equation}
(\Delta G)^2
-\epsilon\bigl[3(\Delta G)^4+\mu_4\bigr]
+O(\epsilon^2).
\label{eq:contrib_01}
\end{equation}

\medskip\noindent
\emph{Pairs $(\alpha,\beta)=(0,j)$ for $j\ge2$}:
Since $|\bra{\psi}G\ket{f_j}|^2=0$ for all $j\ge2$, these terms
are $O(\epsilon^2)$ and do not contribute at first order.

\medskip\noindent
\emph{Pairs $(\alpha,\beta)=(1,j)$ for $j\ge2$}:
These terms are born at $O(\epsilon)$.  With
$\lambda_1=\epsilon(\Delta G)^2$ and $\lambda_j=O(\epsilon^2)$,
\[
\frac{(\lambda_1-\lambda_j)^2}{\lambda_1+\lambda_j}
=\epsilon\,(\Delta G)^2+O(\epsilon^2).
\]
At zeroth order in the matrix elements, the total contribution is
$\epsilon(\Delta G)^2\!\sum_{j\ge2}|\bra{\phi}G\ket{f_j}|^2$.
Evaluating the sum by completeness with the
basis $\{\ket{\psi},\ket{\phi},\ket{f_2},\dots\}$
and using~\eqref{eq:phi}, one finds
\begin{equation}
\sum_{j\ge2}\bigl|\!\bra{\phi}G\ket{f_j}\!\bigr|^2
=
\frac{\mu_4}{(\Delta G)^2}
-(\Delta G)^2
-\frac{\mu_3^2}{(\Delta G)^4}\,,
\label{eq:sum_phi}
\end{equation}
so the $(1,j)$ pairs contribute
\begin{equation}
\epsilon\left[
\mu_4-(\Delta G)^4
-\frac{\mu_3^2}{(\Delta G)^2}
\right]+O(\epsilon^2).
\label{eq:contrib_1j}
\end{equation}

\medskip\noindent
\emph{All remaining pairs}:
Pairs $(j,k)$ with $j,k\ge2$ have both eigenvalues $O(\epsilon^2)$,
yielding an $O(\epsilon^2)$ contribution.\\
Inserting \eqref{eq:contrib_01} and~\eqref{eq:contrib_1j}
into~\eqref{eq:qfi_half}, we find that the QFI to first order in $\epsilon$ is given by
\begin{equation}
\frac{F_\Phi}{4 \nu t^2}
\approx
(\Delta G)^2
+\epsilon\Bigl[
-3(\Delta G)^4-\mu_4
+\mu_4-(\Delta G)^4
             -\frac{\mu_3^2}{(\Delta G)^2}
\Bigr]
\end{equation}
The fourth moment $\mu_4$ cancels between the two contributions, giving
\begin{equation}
F_\Phi
\approx
4 \nu (\Delta G)^2\,t^2
\left\{
1-\epsilon\left[
4(\Delta G)^2
+\frac{\mu_3^2}{(\Delta G)^4}
\right]
\right\},
\label{eq:qfi_shorttime}
\end{equation}
with $\epsilon=\gamma t$,\;
$(\Delta G)^2=\langle G^2\rangle-\langle G\rangle^2$,\; and
$\mu_3=\langle(G-\langle G\rangle)^3\rangle$.
For probe states whose probability distribution over $G$-eigenvalues is symmetric about its mean (\textit{e.g.}\ GHZ or mirror-symmetric superpositions of Dicke states),
$\mu_3=0$ and the result simplifies to
\[
F_\Phi=4 \nu (\Delta G)^2\,t^2\bigl[1-4\epsilon\,(\Delta G)^2\bigr]
+O(\epsilon^2),
\]
where the state-dependent degradation is proportional to \(((\Delta G)^2)^2\), equivalently, for fixed \(\nu\) and \(t\), it is proportional to the square of the noiseless QFI.

\section{Methods}
\label{app:methods}

The master equation in~\eqref{eq:master_eq_correlated} leaves the state confined to the symmetric subspace. 
Restricting ourselves to the symmetric subspace with fixed total spin $J$, the state can be expanded in terms of the $2J+1$ Dicke states $\ketJ{m}$ instead of total Hilbert space dimension $2^{2J}$ for $N=2J$ spins-$1/2$. 
\\
The matrix equation for an individual element of the density matrix reads
\[\label{eq:drhomn}
\hspace{-0.1cm}\dT{t}\rho(t)_{mn} = \rho(t)_{mn} \left[-\im\param(m-n)-\gamma/2 (m-n)^2\right] ,
\]
where $\hat O_{mn}=\braJ{m}\hat O\ketJ{n}$ is the matrix elements of the operator $\hat O$. 
Each term in the $\Jz$ basis evolves independently, as clearly visible in \eqref{eq:drhomn}, so we can easily solve the system for arbitrary time
\[\label{eq:rhonm_correlated_app}
\rho(t)_{mn} = e^{-\im\param t(m-n)-\gamma t/2 (m-n)^2} \rho(0)_{mn}
\]
and the derivative with respect to the parameter $\param$ is given by
\[
\dT{\param}\rho(t)_{mn} = -\im t (m-n) \rho(t)_{mn} .
\]
In this paper, we make use of the explicit form of the dynamics in both analytical and numerical treatments.
\\
In the numerical approach, we compute the QFI directly using
\[
\qfi = 2\sum_{ij: \,\lambda_i+\lambda_j> tol} 
\frac{|\bra{i}\partial_\param\rho_\param\ket{j}|^2}{\lambda_i+\lambda_j}
\]
where $\rho_\param = \sum_j\lambda_j\ket{j}\bra{j}$ is the eigendecomposition of the probe state that we obtain with numerical exact-diagonalization, and we introduce the tolerance $tol=10^{-12}$ to avoid instabilities of the inverse. 
The QFI can be evaluated numerically because the problem reduces to diagonalization within the symmetric subspace, whose dimension scales as $\mathcal O(J)$ and therefore remains numerically manageable even for large values of $J$. 

\noindent The numerical optimization has been employed to determine optimal sensing parameters for the classical and quantum Cram\'er--Rao bounds. 
The numerical procedure to evaluate the optimal parameter and time of the estimation protocol relies on the exact diagonalization. 
There we perform continuous numerical optimization to optimize for the greatest FI and QFI under fixed total time $T$. 
The optimization is taken around the exact optimal parameter and optimal interrogation time to determine the corrections to the known limits. 

\section{Optimal estimator for the SCS probe}
\label{app:scs_optimal_is_Jy}

Let us show that the SLD operator, \textit{i.e.}, optimal estimator, for the probe state $\hat \rho_0 = \ketbra{+}{+}^{\otimes N}$ with $N=2J$ spins, in which we encode the parameter unitarily as $\hat \rho_\param = e^{-i \param \hat{H}} \hat \rho_0 e^{i \param \hat{H}}$ is proportional to $\hat{J}_y$.\\
First, note that under the above unitary encoding, the SLD transforms as $\hat{L}_\param = e^{-i \param \hat{H}} \hat L_0 e^{i \param\hat{H}}$, where according to the definition of the SLD, $\hat L_0$ satisfies
\begin{align}\label{eq:initial_SLD}
    2i\comm{\hat \rho_0}{\hat H} = \{\hat L_0, \hat \rho_0 \}
\end{align}
\\
We evaluate the SLD in~\eqref{eq:initial_SLD} for $\hat \rho_0 = \ketbra{+}{+}^{\otimes N}$ and $\hat H\! = \!\Jz$
\begin{align}
\text{LHS: }\, &2 i \comm{\ketbra{+}{+}^{\otimes N}}{\Jz} \\
\text{RHS: }\, &i  \sum_{k=1}^N \left(\ketbra{+}{+}^{\otimes N} \z{k} - \z{k} \ketbra{+}{+}^{\otimes N} \right)\nonumber\\
&= i \sum_{k=1}^N  \ketbra{+}{+}^{\otimes N-1} \otimes \comm{\ketbra{+}{+}}{\z{k}} ,
\end{align}
For a single spin, it is straightforward to verify that $i\comm{\ketbra{+}{+}}{\hat \sigma_z} = \{\ketbra{+}{+}, \hat \sigma_y \} $, which immediately implies
\begin{align} \label{eq:commJz_anticommJy}
 2i \comm{\ketbra{+}{+}^{\otimes N}}{\Jz} = 2 \{\hat J_y ,  \ketbra{+}{+}^{\otimes N} \}, 
\end{align}
and, comparing with the expression in~\eqref{eq:initial_SLD}, clearly indicates $\hat{L}_0 = 2\hat{J}_y$. 

\section{Spin projection measurement}
\label{app:Jy}

In this section, we calculate the error-propagation sensitivity for the spin projection measurement
\[\label{eq:cfi_Jy}
\cfiEP=\frac{\nu \abs{\partial_\xi \expval{\Jy}_{\xi} }}{\Delta^2 {\Jy}}^2  \eval_{\xi=\param} ,
\]
where $\Delta^2 {\Jy} = \expval{\Jy^2} - \expval{\Jy}^2$.
As discussed below~\eqref{eq:FI_moments}, $\Jy$ is a multi-outcome observable, so~\eqref{eq:cfi_Jy} is the sensitivity of the sample-mean estimator and lower bounds the classical FI of the full $\Jy$ outcome distribution. 
We derive this quantity by moving to the Heisenberg picture of the encoding evolution.
For the encoding model in~\eqref{eq:master_eq_correlated}, the unitary evolution commutes with the noise model, so we can perform the unitary and noise evolution independently. 
\\
First, we calculate the operator $\Jy$ in the Heisenberg picture of noise
\[
e^{\gamma t\mathcal D}\circ\Jy = \sum_{k=0}^\infty \frac{(\gamma t)^k}{k!}{\mathcal D}^k\circ\Jy , 
\]
where for the dephasing we have $\mathcal{D}^\dagger\!=\!\mathcal{D}$. 
The expression can be simplified by observing that $\mathcal{D}[\Jy] = -\frac{1}{2}\Jy$, which yields the following form for the propagated operator in the presence of noise
\[
e^{\gamma t\mathcal D}\circ\Jy = e^{-\gamma t/2} \Jy. 
\]
Unitary evolution can be treated similarly to dephasing noise. 
For the Heisenberg picture, we write
\[
e^{t\param h^{\dagger}}\circ\Jy = \sum_{k=0}^\infty \frac{(\param t)^k}{k!}{h^\dagger}^k\circ\Jy, 
\]
where $h^\dagger[\circ] = \im[\circ, \Jz]$. 
Finally, the $\Jy$ operator in the Heisenberg picture reads
\[
{\Jy}(t) = e^{-\gamma t/2} \left(\sin\param t {\Jx} + \cos\param t  {\Jy}\right) ,
\]
which corresponds to the rotation around the $\hat z$ direction by the angle $\param t$ induced by the unitary.
\\
We now proceed to evaluate $\Jy^2$ in the Heisenberg picture. 
Since the time evolution is not unitary in this setting, we have ${\Jy^2}(t)\neq{\Jy}(t){\Jy}(t)$, and therefore the time-evolved squared operator must be propagated and computed independently. 
We repeat the procedure of calculating the operator in the image of dephasing 
\[
e^{\gamma t \mathcal D}\circ\Jy^2  = \Jy^2 + \frac{1}{2}\left(1-e^{-2\gamma t}\right)(\Jx^2 - \Jy^2) ,
\]
where we used the observation that $\mathcal{D}[\Jy^2] = \Jx^2-\Jy^2$ and $\mathcal{D}[\Jx^2] = -\mathcal{D}[\Jy^2]$. 
We complete the derivation by applying the unitary rotation to the rotated operator
\begin{align}
e^{t\param h^{\dagger}}\circ\Jx^2  &= \left(\cos\param t\Jx-\sin\param t\Jy\right)^2 , \\
e^{t\param h^{\dagger}}\circ\Jy^2  &= \left(\sin\param t\Jx+\cos\param t\Jy\right)^2 . 
\end{align}
Finally, we write the $\Jy^2$ operator in the Heisenberg picture
\begin{align}
\hspace{-.15cm}\Jy^2(t) = \frac{1}{2} &\bigg(\Jx^2f_- + \Jy^2f_+ + \{\Jx,\Jy\}e^{-2\gamma t}\sin2\param t\bigg) ,
\end{align}
where $f_\pm=(1\pm e^{-2\gamma t}\cos2\param t)$. 
\\
In the next step, we collect formulas derived from the previous step to construct the Fisher information in~\eqref{eq:cfi_Jy}. 
We write down the overlaps with the initial state
\begin{align}
\expval{\Jy}\!(t) = e^{-\gamma t/2} \left(\sin\param t \expval{\Jx} + \cos\param t \expval{\Jy}\right) ,
\end{align}
and
\begin{align}
\hspace{-.12cm}\expval{\Jy^2}\!(t) \!= \! \frac{\expval{\Jx^2} f_- \!+\! \expval{\Jy^2} f_+ \!+\! 2\Re\expval{\Jx\Jy} e^{-2\gamma t}\sin2\param t}{2}, 
\end{align}
where $\expval{\hat O}$ is calculated for the input probe. 
Similarly, the derivative reads 
\begin{align}
\hspace{-0.18cm}\dT{\param}\expval{\Jy}\!(t) \!=\! e^{-\gamma t/2} t \left(\cos\param t \expval{\Jx} \!- \sin\param t \expval{\Jy} \right) .
\end{align}
\\
The ansatz of GSS in~\eqref{eq:probestate_definition} is symmetric under reflection in the $xz$ and $xy$ planes. 
The former implies $\expval{\Jy}=0$ and $\expval{\Jx\Jy}=0$ and the expression simplifies to
\begin{align}
\cfiEP = \frac{\nu t^2 {2e^{-\gamma t} \cos^2\param t \expval{\Jx}^2} }{\expval{\Jx^2} f_-+\Delta^2\Jy f_+ -  2e^{-\gamma t}\sin^2\param t \expval{\Jx}^2},
\end{align}
The optimal parameter that maximizes $\cfiEP$ is $\paramOpt=n \pi/t$ with integer $n$.
The expression for the optimal parameter
reads
\begin{align}
\cfiEP =& \frac{2 \nu e^{-\gamma t} t^2 \expval{\Jx}^2}{\expval{\Jx^2}(1-e^{-2\gamma t})+\Delta^2\Jy(1+e^{-2\gamma t})}. 
\end{align}
In the static spin approximation, where the observables $\expval{\Jx}\approx J$ and $\expval{\Jx^2}\approx J^2$ for a weak squeezing, the expression becomes a function of $\Delta^2\Jy$.
There, the sensitivity increases as the variance $\Delta^2\Jy$ decreases.
This indicates that the squeezing improves the precision of estimation. 
The optimal time for the squeezed state is shifted from the one that is optimal for the classical state and reads 
\[\label{eq:topt_Jy}
t_{\rm opt} \overset{\paramOpt}{\approx}  \frac{1}{\gamma}\left(\frac{3\Delta^2\Jy}{J^2}\right)^{1/3} ,
\]
for weak squeezing $\Delta^2\Jy\ll J^2$ and $\gamma t\ll 1$. 
The optimized Cram\'er--Rao bound reads
\begin{equation} \label{eq:GSSopt_CRB}
\Delta^2 \paramOpt
\overset{\tOpt}{\gtrsim}
\frac{\gamma}{T}
\left[
1
+ 
\frac{1}{2}
\left(
\frac{3\Delta^2\Jy}{J^2}
\right)^{\!\!2/3}
\right] ,
\end{equation}
where the bound is valid for weak squeezing and $J\gg 1$ and for $\gamma \tOpt\ll1$. 
Notice that the convergence in the function of $J$ depends on the scaling of the variance of the squeezed dimension. 
For the spin coherent state, where $\Delta^2\Jy=J/2$, we can verify that $\Delta^2 \paramOpt\overset{\tOpt}{\gtrsim}\frac{\gamma}{T}\left[1+ \frac{1}{2}\left(\frac{3}{2J}\right)^{2/3}\right]$.  
Similarly, for the optimal state in two axis twisting~\cite{kitagawa_squeezed_1993}, where $\Delta^2\Jy\sim \frac{1}{2}$, we can show that 
$\Delta^2 \paramOpt\overset{\tOpt}{\gtrsim}\frac{\gamma}{T}\left[1+ \frac{1}{2}\left(\frac{3}{2J^2}\right)^{2/3}\right]$. 
From the analysis, we can conclude that the squeezing improves the estimation only by a subleading factor, and the estimation precision is ultimately always limited from below by $\gamma/T$ set by the noise amplitude.

\subsection{Spin coherent state}
\label{app:scs_Jy}

In this section, we derive the Fisher information for the spin coherent state. In addition, we comment on the optimal conditions and the Cram\'er--Rao bound for the estimation. 
In the Dicke basis $\ket{\pi/2, \sigma_{\rm SCS}}$ is given by
\begin{align}
    \ket{\pi/2, \sigma_{\rm SCS}} = \frac{1}{2^{J}}\sum_{m=-J}^J \sqrt{\binom{2J}{J+m} }\ketJ{m} .
\end{align}
Now, we use \eqref{eq:cfi_Jy} to compute the achievable estimation precision
\begin{align}\label{app:CRLB_Jy_sep_corr} 
\Delta^2 \param \geq \frac{(1 + 2J)e^{2\gamma t} + (1 - 2J)\cos(2\param t) - 4J e^{\gamma t}\sin^2(\param t)}
{4J T t\, e^{\gamma t}\cos^2(\param t)} .
\end{align}
We determine the locations of the extrema of the Fisher information in~\eqref{app:CRLB_Jy_sep_corr} with respect to the parameter and thereby obtain the optimal parameter value as $\paramOpt = n \pi/t$, where $n$ is an integer. 
The Cram\'er--Rao bound for the optimal parameter reads
\begin{align} \label{eq:CRBSCSquasioptimal}
\Delta^2 \paramOpt &\geq \frac{1}{4 J T t} \left[e^{-\gamma t}(1 - 2J) + e^{\gamma t}(1 + 2J) \right] .
\end{align}
The expression can be rewritten as
\[
\text{RHS: }\quad \frac{\gamma}{T} \left( \frac{\frac{1}{2J}\cosh x + \sinh x}{x} \right) \equiv \frac{\gamma}{T} \, h(x) 
\]
where in the last equality, we introduced $h(x)$ with the argument $x=\gamma t$. 
Now our goal is to minimize $h(x)$ over all $x>0$. 
We compute the derivative of the function and set it equal to zero, which yields the necessary condition
\begin{align} \label{eq:minimumaux1}
\frac{x\sinh x - \cosh x}{2J} + x\cosh x - \sinh x = 0 . 
\end{align}
In order to solve the equation, we expand the expression for $x \ll 1$ giving
\begin{align}\label{eq:xs_expressions_approx}
&x \cosh x - \sinh x \simeq \frac{x^3}{3} + \mathcal{O}(x^5)  ,\\
&\frac{1}{2J}\left( x \sinh x - \cosh x \right)
\simeq -\frac{1}{2J} + \frac{x^2}{4J} + \mathcal{O}(x^4) ,
\end{align}
where we kept the leading terms in terms of the argument $x=\gamma t$. 
Substituting the approximate formulas in~\eqref{eq:xs_expressions_approx} into \eqref{eq:minimumaux1}, and applying the limit of large $J\gg 1$, we obtain the optimal argument $x=x_{\rm opt}$. 
From this we calculate the optimal interrogation time $\tOpt=x_{\rm opt}/\gamma$
\begin{align}
    \tOpt \overset{\paramOpt}{\approx}\frac{1}{\gamma} \left( \frac{3}{2J} \right)^{1/3} .
\end{align}
\\
Now, we can compute the Cram\'er--Rao bound in~\eqref{eq:CRBSCSquasioptimal} for the optimal parameters. 
In the weak noise and short term approximation, such that $\gamma t\ll1$, the exponent can be simplified, and we arrive at the expression for the ultimate precision for the spin-projection readout
\[\label{eq:d2Phi_Jy_paramOpt_tOpt_scs}
\Delta^2\paramOpt \overset{\tOpt}{\gtrsim} \frac{\gamma}{T}\left[1+\frac{1}{2}\left(\frac{3}{2J}\right)^{2/3}\right] ,
\]
which for large systems is lower bounded by $\Delta^2\paramOpt \overset{\tOpt}{\geq}\frac{\gamma}{T}$ and the subleading correction converges as $J^{-2/3}$ with the system size. 

\subsection{Squeezed state}
\label{app:sss_Jy}

In this section, we derive the bound on the estimation precision for the squeezed state modeled by the GSS ansatz in~\eqref{eq:probestate_definition}. 
For the squeezed state, we can calculate the observables by taking the large system size limit and going to the continuous limit. 
There, the observables reduce to Gaussian integrals, which are easier to evaluate than a discrete sum. 
In order to derive the bound, we rely on a general formula in~\eqref{eq:GSSopt_CRB}. 
\\
First, let us derive the expression for the transverse variance $\Delta^2\Jy$ which will be related to the squeezing parameter of the state
From the symmetry of the state we know that $\expval{\Jy}=0$. 
For the $\expval{\Jy^2}$, we write the expectation value 
\begin{widetext}
\begin{align}
\expval{\Jy^2} = \frac{1}{4} 
\sum_{m,m'} \overline{c_m}{c_{m'}} \bigg[&\delta_{m,m'} (J(J+1) - 2 m')-\delta_{m,m'+2}\sqrt{J(J+1)-m(m+1)}\sqrt{J(J+1)-(m+1)(m+2)}\nonumber\\
&-\delta_{m,m'-2}\sqrt{J(J+1)-m(m-1)}\sqrt{J(J+1)-(m-1)(m-2)} \bigg] ,
\end{align} 
\end{widetext}
where $c_m$ give normalize state coefficients and the range of the summation takes $m'+2\leq J$ and $m'-2\geq -J$ for the second and third terms, respectively. 
After some algebra and collecting terms, we obtain
\begin{widetext}
\begin{align}
\label{eq:Jy2}
    \expval{\Jy^2}
    &= \frac{1}{2}J(J+1)
    - \frac{1}{2}\expval{\Jz^2} \nonumber\\
    &\quad
    - \frac{1}{2}
    \sum_{m=-J}^{J-2}
    c_m c_{m+2}
    \sqrt{J(J+1)-m(m+1)}
    \sqrt{J(J+1)-(m+1)(m+2)}.
\end{align}
\end{widetext}

In the large-system limit, \textit{i.e.}, $J\gg 1$, the variance along the
$z$ direction is well approximated by
\[
\expval{\Jz^2} \approx \sigma^2.
\]

We now turn to the last term in~\eqref{eq:Jy2}. Introducing the shifted
variable
\[
u=m+1,
\]
the relevant combinations of $m$ become
\begin{align*}
m^2+(m+2)^2 &= 2u^2+2,\\
m(m+1) &= u(u-1),\\
(m+1)(m+2) &= u(u+1).
\end{align*}

Using these relations, the last term in Eq.~\eqref{eq:Jy2} can be written as
\begin{equation}
\label{eq:Jy2_last_term}
-e^{-1/(2\sigma^2)}
\frac{1}{2}
\sum_{m=-J}^{J-2}
p(u)\sqrt{X^2-u^2},
\end{equation}
where
\[
p(u)
=
\normN{\ket{\Psi(\pi/2,\sigma)}}^2
e^{u^2/(2\sigma^2)},
\]
and
\[
X=J(J+1)-u^2.
\]

For the squeezed state, the distribution is concentrated around values of
$u$ that are much smaller than the characteristic scale set by $X$.
Therefore, to leading order, we approximate
\[
\sqrt{X^2-u^2}\approx X.
\]
Replacing the sum by its continuum approximation then gives
\[
-e^{-1/(2\sigma^2)}
\left[J(J+1)-\sigma^2\right].
\]
Substituting this result, together with
$\expval{\Jz^2}\approx\sigma^2$, into~\eqref{eq:Jy2}, we finally obtain
\[
\expval{\Jy^2}
=
\frac{1}{2}
\left[J(J+1)-\sigma^2\right]
\left(1-e^{-1/(2\sigma^2)}\right).
\]

Working in the limit of $1\ll \sigma \ll J$, we can approximate the expression by the $\Delta^2{\Jy} \approx \frac{1}{4} \left(\frac{J}{\sigma}\right)^2$. 
Inserting the variance into~\eqref{eq:GSSopt_CRB}, we get the Cram\'er--Rao bound for the squeezed state
\begin{align}
\label{eq:d2Phi_Jy_paramOpt_tOpt_sss}
\Delta^2\paramOpt\overset{\tOpt}{\geq}& \frac{\gamma}{T}\left[1+\frac{1}{2}\left(\frac{3}{4}\right)^{2/3}\!\!\!\sigma^{-4/3}\right].
\end{align}
We can verify that the analytical expression in~\eqref{eq:d2Phi_Jy_paramOpt_tOpt_sss} reduces to the result calculated for SCS in~\eqref{eq:d2Phi_Jy_paramOpt_tOpt_scs} when we use $\sigma^2=J/2$. 
The expression in~\eqref{eq:d2Phi_Jy_paramOpt_tOpt_sss} reveals that the estimation precision improves with the squeezing parameter controlled by the variance $\sigma$. 

\begin{figure*}[t!]
    \centering
    \includegraphics[width=\columnwidth]{./fig7.pdf}
    \includegraphics[width=\columnwidth]{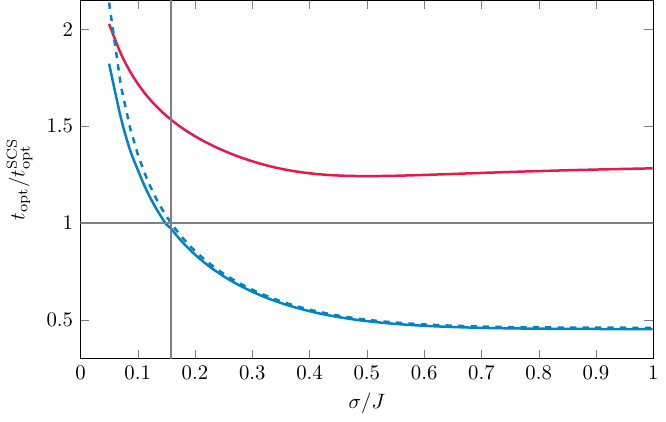}
    \caption{
    {\bf The optimal error-propagation and quantum Cram\'er--Rao bounds for spin-squeezed states. }
    The error-propagation Cram\'er--Rao bound (blue line) and the quantum Cram\'er--Rao bound  (red line) in panel (a) was calculated for the optimal interrogation time presented in the other panel.  (b) The optimal time is measured as the fraction of the time in~\eqref{eq:topt_Jy}. 
    Data obtained at the optimal parameter $\paramOpt=n\pi/t$ for the system size with $J=20$ under correlated dephasing with $\gamma=0.01$. 
    Gray lines are included as visual guides. The vertical line indicates the spin coherent state. 
    }
    \label{fig:topt_SCS_like}
\end{figure*}

\noindent In Figure~\ref{fig:topt_SCS_like} we numerically verify the analytical expressions by plotting them against the numerically-obtained optimum bound. 
There, we also contrast the bound calculated for the spin projection with the quantum Cram\'er--Rao bound set by the QFI.
For the details on the numerical optimization strategy, see App.~\ref{app:methods}. 
The resulting optimized QFI and $\cfiEP$ bounds for the squeezed state are shown in Fig.~\ref{fig:topt_SCS_like}(a).
We demonstrate that, although the optimal time maximizing $\cfiEP$ is consistent with the analytical expression given in~\eqref{eq:topt_Jy}, the interrogation time that optimizes the overall strategy is shifted to larger values, see Fig.~\ref{fig:topt_SCS_like}(b).
Similarly, the quantum Cram\'er--Rao bound provides a tighter limit than its spin-projection counterpart, as illustrated in Fig.~\ref{fig:topt_SCS_like}(a), thereby resulting in an overall enhancement of the phase-estimation sensitivity.
This improvement is to be expected, since the QFI upper bounds the classical FI of any measurement, which in turn upper bounds the moment-based sensitivity $\cfiEP$.

\section{Parity measurement}
\label{app:Par}

In this section, we calculate the Fisher information for the parity measurement
\[\label{eq:cfi_Par}
\cfi=\frac{\nu \abs{\partial_\xi \expval{\Par}_{\xi} }}{1 -  \expval{\Par}^2}^2  \eval_{\xi=\param} ,
\]
where we used the fact that the variance of the parity operator reads $\Delta^2 {\Par} = 1-\expval{\Par}^2$ where we used that $\Par^2=\mathbbm 1$ and the parity operator is $\Par =\prod_{j=1}^N\x{j}$, where $N=2J$ is the number of spins. 
In the Dicke basis, the operator reads
\[
\Par = \sum_{m=-J}^J \ketJ{-m}\braJ{m} ,
\]
which is equivalent to flipping orientation of all spins in $\ketJ{m}$ superposition. 
\\
In the next steps, we will use the same approach as in App.~\ref{app:Jy} and calculate the expectation values by moving to the Heisenberg picture of the master equation in~\eqref{eq:master_eq_correlated}. 
Since unitary evolution of the encoding commutes with the noise model, we can perform these steps sequentially. 
We begin the derivation by observing that the noise and encoding generators transform the parity operator in the following manner
\begin{align}
&\mathcal{D}^\dagger[\Par] = -2\Jz^2\Par \text{, and} \quad {h}^\dagger[\Par] = 2\im\Jz \Par .
\end{align}
As a consequence and as a fact that $\Jz$ commutes with both propagators, we can write the parity operator in the Heisenberg picture
\[
\Par(t) = \hat \Pi(t) \Par , 
\]
where we define auxiliary operator carrying time-dependence
\[
\hat \Pi(t) = e^{-2\gamma t \Jz^2 + 2\im\param t \Jz} .
\]
The operator is diagonal in the Dicke basis
\[
\hat \Pi(t) = \sum_{m=-J}^J e^{-2\gamma t m^2 + 2\im\param t m}\ketJ{m}\braJ{m} .
\]
For states with mirror symmetry with respect to the $xy$ plane, that is,  $c_m=c_{-m}$, the parity operator $\Par\ket{\psi} = \ket{\psi}$, therefore, Fisher information for noisy sensing model simplifies to 
\[\label{eq:cfi_Par_XYmirror}
\cfi=\frac{\nu \abs{\partial_\xi \expval{\hat \Pi(t)}_{\xi} }}{1 -  \expval{\hat \Pi(t)}^2}^2  \eval_{\xi=\param} . 
\]
Writing the expectation values as sums, we arrive at the formula
\[\label{eq:cfi_parity}
\cfi = \frac{4tT\left|\sum_m|c_m|^2 m e^{-2\gamma t m^2}\sin(2\param t m)\right|^2}
{1 - \left(\sum_{m}|c_m|^2 e^{-2\gamma t m^2}\cos(2\param t m)\right)^2} ,
\]
for normalized distribution $c_m$. 
By analyzing the expression, it is difficult to deduce the optimal parameter $\paramOpt$ since the encoding dependence is a function of the momenta $m$. 
Note that, in contrast to the spin-projection measurement, the dependence on the relevant parameters cannot be readily factorized. 
In the next subsections, we will consider specific cases of the input state, in particular the GHZ states and GHZ-like states. 

\subsection{GHZ state}
\label{app:ghz_Par}

\noindent In the Dicke basis, the GHZ state is written as
\begin{align} \label{eq:GHZprobeDicke}
\ket{GHZ} = \frac{1}{\sqrt2} \left(\ketJ{J} + \ketJ{-J}\right).   
\end{align}
The density matrix of the GHZ state after time $t$ reads
\begin{align}
\rho (t)= \frac{1}{2} \bigg(&\ketbra{J,J}{J,J}  + \ketbra{J,-J}{J,-J} \nonumber\\
&+ e^{- 2 i\param tJ-2 J^2 \gamma t} \ketbra{J,J}{J,-J} \nonumber\\
&+ e^{2 i\param t J-2 J^2 \gamma t}\ketbra{J,-J}{J,J}\bigg) . 
\end{align}
for which the corresponding Fisher information can be readily computed
\begin{align}
\hspace{-0.3cm}\Delta^2 \param \geq\! \frac{e^{4 \gamma  J^2 t} \csc ^2(2 \param  J t) \left[1\!-\!e^{-4 \gamma  J^2 t} \cos ^2(2 \param  J t)\right]}{4 J^2 t T} .
\end{align}
The Cram\'er--Rao bound is minimized for the optimal parameter $\paramOpt= \frac{\pi}{4 J t}$, giving a precision bound
\begin{align}
    \Delta^2 \paramOpt \geq \frac{1}{e^{-4 \gamma t J^2} 4 J^2 t T} .
\end{align}
It is straightforward to find the exact $t$ for which the minimum occur which is
\[\label{eq:topt_GHZ}
\tOpt \overset{\paramOpt}{=} \frac{1}{4 \gamma J^2}. 
\]
Substituting the optimal time, we arrive at the ultimate precision bound
\[
\Delta^2 \paramOpt \overset{\tOpt}{\geq}  \frac{e \gamma}{T} , 
\]
which is determined by the noise strength and is independent of the system size. 

\subsection{GHZ-like state}
\label{app:ghz_like_par}

We begin the derivation from the general formula in~\eqref{eq:cfi_parity}. 
To simplify the formula we take the sums
\begin{align}
{M}_0 &= \sum_{m}|c_m|^2 e^{-2\gamma t m^2}\cos(2\param t m)\\
{M}_1 &= \sum_m|c_m|^2 m e^{-2\gamma t m^2}\sin(2\param t m)
\end{align}
and take their corresponding continuous limits
\begin{align}
{\cal M}_0 &= 
J \int_{-1}^1 dx \, |c_x|^2 e^{-2\gamma t J^2 x^2} \cos(2\param t J x)\\
{\cal M}_1 &= 
J^2 \int_{-1}^1 dx \, |c_x|^2 e^{-2\gamma t J^2 x^2} \sin(2\param t J x)\\
\end{align}
where $m\rightarrow x J$ and $c_m\rightarrow c_x$. 
The coefficient amplitude for the GHZ-like states are 
\begin{align}
|c_x|^2 
=&  \normN{\catTh{0,\sigma}}^2 \left(e^{-\frac{J^2}{4\sigma^2}(x-1)^2} + e^{-\frac{J^2}{4\sigma^2}(x+1)^2}\right)^2\\
=&  \normN{\catTh{0,\sigma}}^2 \bigg(e^{-\frac{J^2}{2\sigma^2}(x-1)^2} + e^{-\frac{J^2}{2\sigma^2}(x+1)^2}
+ e^{-\frac{J^2}{2\sigma^2}(x^2+1)}
 \bigg) \label{eq:cx_ghzlike_expand}\\
\approx&  \frac{1}{\sigma\sqrt{2\pi}}\bigg(e^{-\frac{J^2}{2\sigma^2}(x-1)^2} + e^{-\frac{J^2}{2\sigma^2}(x+1)^2} \bigg) \label{eq:cx_ghzlike_simplified}
, 
\end{align}
where we used the amplitude 
\[
c_x =  \normN{\catTh{0,\sigma}}  \left(e^{-\frac{J^2}{4\sigma^2}(x-1)^2} + e^{-\frac{J^2}{4\sigma^2}(x+1)^2}\right)
\]
and the expression in~\eqref{eq:cx_ghzlike_expand} is simplified to the expression in~\eqref{eq:cx_ghzlike_simplified} by dropping the coherence term that is exponentially suppressed for $\sigma\ll J$. 
The integrals become 
\begin{align}
{\cal M}_0 &\approx 
\frac{2J}{\sigma\sqrt{2\pi}} \int_{-1}^1 dx \, e^{-\frac{J^2}{2\sigma^2}(x-1)^2 -2\gamma t J^2 x^2} \cos(2\param t J x) , \label{eq:M0}\\
{\cal M}_1 &\approx  
\frac{2J^2}{\sigma\sqrt{2\pi}} \int_{-1}^1 dx \, x\, e^{-\frac{J^2}{2\sigma^2}(x-1)^2-2\gamma t J^2 x^2} \sin(2\param t J x). \label{eq:M1}
\end{align}

The exponential decay $e^{-\frac{J^2}{2\sigma^2}(x-1)^2}$ strongly localizes the exponent around $x=1$, therefore we approximate the other functions under the Taylor expansion up to the quadratic terms of $x-1$. 
\[
f_k(x)\bigg|_{x=1}\approx f_k(1) + (x-1) f'_k(1) + (x-1)^2\frac{f''_k(1)}{2} ,
\]
where the derivative is taken with respect to $x$ and 
\begin{align}
f_0(x) &=  e^{-2\gamma t J^2 x^2} \cos(2\param t J x) , \\
f_1(x) &= x\, e^{-2\gamma t J^2 x^2} \sin(2\param t J x).
\end{align}
With the approximation above, the expressions in~\eqref{eq:M0} and~\eqref{eq:M1} become just a simple Gaussian integrals
\begin{align}
{\cal K}_0 =& \int_{-1}^1 dx \,  e^{-\frac{J^2}{2\sigma^2}(x-1)^2}  = \sqrt\frac{\pi}{2} \epsilon\,  {\rm erf}(\frac{\sqrt2}{\epsilon})  \\\approx& \sqrt\frac{\pi}{2} \epsilon ,\label{eq:K0}\\
{\cal K}_1 =& \int_{-1}^1 dx \,  e^{-\frac{J^2}{2\sigma^2}(x-1)^2} (x-1) = \epsilon^2 \left(e^{-2/\epsilon^2}\!-\!1\right)  \\ \approx& -\epsilon^2 ,\label{eq:K1}\\
{\cal K}_2 =& \int_{-1}^1 dx \,  e^{-\frac{J^2}{2\sigma^2}(x-1)^2} (x-1)^2  = -2 \epsilon^2 e^{-2/\epsilon^2}  \\ &+ \sqrt\frac{\pi}{2}\epsilon^3 \, {\rm erf}(\frac{\sqrt2}{\epsilon}) \approx \sqrt\frac{\pi}{2}\epsilon^3 ,\label{eq:K2}
\end{align}
where $\epsilon=\sigma/J$ in the last step we invoke the approximation for $\epsilon\ll1$. 
Notice, that the amplitude of the integral ${\cal K}_n\propto \epsilon^n$ progressively decreasing its impact on the FI expression. 

\begin{figure*}[t!]
    \centering
    \includegraphics[width=.68\columnwidth]{./fig8.pdf}
    \includegraphics[width=.68\columnwidth]{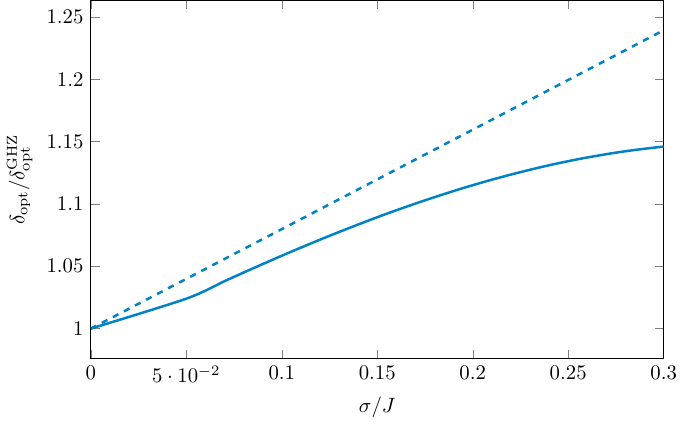}
    \includegraphics[width=.68\columnwidth]{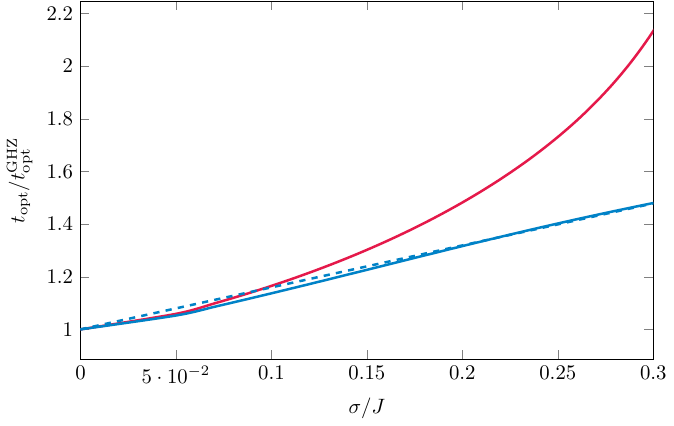}
    \caption{
    {\bf The optimal classical and quantum Cram\'er--Rao bounds for GHZ-like states. }
    (a) The classical Cram\'er--Rao bound  (blue line) and the quantum Cram\'er--Rao bound  (red line) in panel was calculated for the optimal parameter (see panel b) and the optimal interrogation time (see panel c). 
    The numerical values are compared with the analytical expressions (dashed blue line). 
    Data obtained for $J=20$ under correlated dephasing with $\gamma=0.01$. 
    }
    \label{fig:topt_GHZ_like}
\end{figure*}

In order to determine the optimal parameter for the GHZ-like states, we inspect the extremum of the Fisher information. 
To simplify the evaluation we rely on the quadratic approximation of the FI around $\paramOpt^{\rm GHZ} = \frac{\pi}{4 J t}$. 
From the zeroing of the derivative, we find that the optimal parameter is approximately
\[
\paramOpt \approx \frac{\pi}{4 J t}\left(1 + \sqrt\frac{2}{\pi}\epsilon\right). 
\]
Assuming the optimal parameter, we repeat the procedure to find the optimal interrogation time
\[
\tOpt \overset{\paramOpt}{\approx} \frac{1}{4 J^2\gamma}\left(1 + 2\sqrt\frac{2}{\pi}\epsilon\right). 
\]
Finally, we arrive at the optimal parity Fisher information 
\begin{align}
\cfi \overset{\paramOpt, \tOpt}{\approx}& \frac{T}{e\gamma}\left(1- \frac{(\pi-2)(8+\pi^2)}{4\pi} \epsilon^2\right) \\\approx& \frac{T}{e\gamma}\left(1- 1.6234\, \epsilon^2\right) 
\end{align}
and the corresponding Cram\'er--Rao bound
\begin{align}
\Delta^2\paramOpt \overset{\tOpt}{\approx}& \frac{e\gamma}{T}\left(1+ \frac{(\pi-2)(8+\pi^2)}{4\pi} \epsilon^2\right) \\\approx& \frac{e\gamma}{T}\left(1+ 1.6234\, \epsilon^2\right) ,
\end{align}
admitting quadratic correction due to broadening of the GHZ-like distribution. 

\noindent In Fig.~\ref{fig:topt_GHZ_like} we compare the analytic results with the numerically obtained optimal parameters. 
The analytical results qualitatively agree with weakly broadened GHZ-like state. 
We note that the approximation requires both $\sigma/J\ll1$ and $J\gg1$ in order to get the continuum, therefore, for very small $\sigma$ the continuum approximation might break down. 
As the result, the numerically optimized parameters do not fully align with the analytics. 
Nevertheless, we obtain a good approximation of the FI for small $\sigma/J$. 

\section{Precision bound for superposition of Dicke states}
\label{app:Dicke_cat_qfi_fiPar}

The superposition of Dicke states employed in the GSS ansatz in~\eqref{eq:probestate_definition} is given by
\[\label{eq:Jmcat}
\catTh{\Theta, 0} = \frac{1}{\sqrt 2}\left(\ketJ{\mAv(\Theta)}+\ketJ{-\mAv(\Theta)}\right) ,
\]
where $\mAv(\Theta) = J \cos\Theta$ with the angle restricted to $\Theta\in[0,\pi/2]$. 
Each term in the superposition is the eigenstate of $\Jz$, therefore, it has zero variance for $\Jz$. 
Under the master equation in~\eqref{eq:master_eq_correlated}, the corresponding density matrix contains at most four non-vanishing elements
\begin{align}
&\rho(t)_{k,k} = \rho(t)_{-k,-k} = \frac{1}{2} ,\\
&\rho(t)_{k,-k} = \overline{\rho(t)_{-k,k}} = \frac{1}{2} e^{-2 t k(i\param+\gamma k)} ,
\end{align}
where, for the sake of notational simplicity, we have introduced the shorthand $k=\mAv(\Theta)$. 
Similarly, the state derivative has only two non-zero terms on its off-diagonal blocks 
\begin{align*}
\dT{\param}\rho(t)_{k,-k} &=
\dT{\param}\rho(t)_{-k,k}^* = -it k e^{-2 t k(i\param+\gamma k)} .
\end{align*}
For the superposition, we can calculate the sensitivity and degradation indicators in~\eqref{eq:F0_correlated} and in~\eqref{eq:F1_correlated}, respectively. 
The expression depends on the spin orientation and reads exactly
\[
\qfi^{(0)} = 4 \nu t^2 k^2, \quad \qfi^{(1)} = 16 \nu t^2 k^4. 
\]
Notice that the QFI indicates that only the state with $k=0$ is not subject to decoherence, although it is not metrologically useful since its QFI is zero for any system size. 
\\
For the superposition of Dicke states, the sparse matrix form enables easy QFI calculation under noisy encoding
\begin{align}\label{eq:qfiJm_phi_k}
\qfi\left[ \Psi(\Theta, 0)\right] = 4 Tt k^2 e^{-4 k^2 \gamma t},    
\end{align}
where replacing $k=J\cos\Theta$ we recover the angular dependence
\begin{align}\label{eq:qfiJm_phi}
\qfi\left[ \Psi(\Theta, 0)\right] = 4 Tt J^2\cos^2\Theta \,e^{-4 J^2\cos^2\Theta \gamma t}.
\end{align}
For the noiseless scenario, that is, $\gamma=0$, the expression in~\eqref{eq:qfiJm_phi} recovers the Heisenberg limit scaling for any $\Theta$. 
However, the noise introduces exponential suppression that results in an inevitable turnover in QFI. 

\noindent For the Dicke state superposition, the parity measurement introduced in Sec.~\ref{sec:Par} saturates the quantum Cram\'er--Rao bound. 
In the classical estimation protocol we find the optimal parameter
\[
\paramOpt=\frac{\pi}{4kt} ,
\]
where the FI is equal to the QFI in~\eqref{eq:qfiJm_phi_k} and the expression is obtained for $k\neq0$, that is, for  $\Theta\neq\pi/2$
The associated optimal interrogation time for QFI and FI at the optimal parameter reads 
\[
\tOpt=\frac{1}{4\gamma k^2},
\]
Finally, we calculate the ultimate estimation precision
\begin{align} \label{eq:GHZopt_CRB}
    \Delta^2 \paramOpt \overset{\tOpt}{\geq} \frac{e \gamma}{T} ,
\end{align}
which is independent of the system size and the orientation of the Dicke states. 
The bound indicates that any orientation $\Theta$ is metrologically equivalent to the GHZ state.

\end{document}